\documentstyle[aps,pra,epsf]{revtex}
\begin{document}
\draft

\newcommand{\pp}[1]{\phantom{#1}}
\newcommand{\be}{\begin{eqnarray}}
\newcommand{\ee}{\end{eqnarray}}
\newcommand{\ve}{\varepsilon}
\newcommand{\vs}{\varsigma}
\newcommand{\Tr}{{\,\rm Tr\,}}
\newcommand{\pol}{\frac{1}{2}}

\title{
(Noncanonical) field quantization by means of a single harmonic oscillator
}
\author{Marek~Czachor}
\address{
Katedra Fizyki Teoretycznej i Metod Matematycznych\\
Politechnika Gda\'{n}ska,
ul. Narutowicza 11/12, 80-952 Gda\'{n}sk, Poland\\
and\\
Arnold Sommerferld Instit\"ut f\"ur Mathematische Physik\\
Technische Universit\"at Clausthal, 38678 Clausthal-Zellerfeld, Germany}
\maketitle

\begin{abstract}
A new scheme of field quantization is proposed. Instead of
associating with different frequencies different oscillators we begin
with a single oscillator that can exist in a (quantum) superposition of
different frequencies. 
The idea is applied to the electromagnetic radiation field and 
nonrelativistic quantum optics. 
Employing a Dirac-type mode-quantization of the electromagnetic
field and using a single oscillator 
we obtain several standard properties such as coherent states
or spontaneous and stimulated emission. Extending the formalism to a
greater number of oscillators we arrive at a structure analogous to
the Fock space but without the standard cyclic ``vacuum state". In
the modified formalism the notion of the vacuum state is replaced 
by a vacuum {\it subspace\/} spanned by ground states of the
oscillators. As opposed to the
standard approach the vacuum energy is finite and does not have to be
removed by any ad hoc procedure. Atom-light interactions are
described by an appropriately modified minimal-coupling Hamiltonian
(no normal ordering of the free-field Hamiltonian is necessary). 
The Hamiltonian does not change the number of oscillators which leads
to an additional conservation law. 
Using the ``$-(e/m)\vec A\cdot\vec p\,$" interaction we discuss in
second-order perturbation theory a
two-photon spontaneous emission. The result essentially agrees with
the ordinary formulas but the nontrivial vacuum structure is
explicitly seen in the two-photon amplitude. The probability of the
2-photon emission resulting from the new formalism consists of a product 
of several terms, a part of them resembling those arising in the standard
formulation from detector
inefficiency, and the remaining 
one being the well known quantum optics formula. 
The presence of the additional conservation law shows that the theory
contains two kinds of bosons (oscillators, whose number is conserved,
and their excitations, whose number is not conserved). Taking this
distinction into account we calculate an analog of the blackbody
radiation Planck law. For temperatures lower than some $T_{\rm
critical}$  the result is indistinguishable from the
Planck distribution. For $T>T_{\rm critical}$ the distribution is
Planck-like but with the maximum lowered and shifted towards higher
frequencies. 
\end{abstract}


\section{Introduction}

The standard quantization of a harmonic oscillator is based on 
quantization of $p$ and $q$ but $\omega$ is a parameter. To have,
say, two different frequencies one has to consider two independent 
oscillators. On the other hand, it is evident that there can exist
oscillators which are in a {\it quantum superposition\/} of different
frequencies. The example is an oscillator wave packet associated with
distribution of center-of-mass momenta. It is known that the superposition of 
momenta gets translated into a superposition od Doppler shifts 
and therefore also of frequencies. We stress here the word ``quantum"
since the superpositions we have in mind are not those we know from
{\it classical\/} oscillations. 

This trivial observation raises the question of the role of
superpositions of frequencies for a description of a single harmonic
oscillator. The motivation behind the problem is associated with the question
of field quantization: Is it possible that a quantum field consists 
of oscillators whose frequencies are {\it indefinite\/}? If so, maybe to
quantize the field it is sufficient to use only one oscillator which exists 
in a {\it quantum\/} superposition of all the possible frequencies allowed 
by the boundary conditions of a given problem? 

The idea is very simple. It is known that a ``one-particle'' 
state vector can be regarded as a representation of an ensemble of particles 
in a given pure state. On the other hand, the classical electromagnetic 
field can be regarded as an ensemble of oscillators. The standard idea of 
quantization, going back to 1925 \cite{BHJ}, 
is to treat the field as an ensemble of 
quantum oscillators. But the ensemble itself is, in a sense, a classical one 
since for each frequency we need a separate oscillator. This is analogous 
to a classical ensemble of particles forming a classical wave on a lake 
surface.
For each point on the surface we need a separate particle because a
classical particle can ocupy only a single point in space. A quantum wave 
is of course different and we are all accustomed to the idea of a 
single-particle wave. In this case the properties of the entire ensemble are 
somehow encoded in properties of a single element of the ensemble. 

For some reasons, probably partly historical and sociological, it seems 
that the idea of 
a single-particle state vector representation of the ensemble of oscillators
has never been considered. 
The historical reason may be the fact that the very concept of field 
quantization occured already in 1925. At that stage quantum mechanics existed 
still in a matrix form and the Schr\"odinger paper ``Quantisierung als 
Eigenwertproblem''\cite{Sch}, where the Schr\"odinger equation occured for the
first time and the role of eigenvalues was explained, was not yet published. 
Sociologically, the names and reputation of Heisenberg, Born, Jordan, Dirac,
together with the unquestionable success of quantum optics, field theory, and 
statistical physics, made it almost impossible to question the very 
starting point of the theory. The ideas presented below are an accidental 
by-product of a work on a different problem. 

It should be mentioned that several approaches towards an alternative 
description of the electromagnetic field at a fundamental level were 
proposed 
(e.g. Janes' \cite{Janes} 
neoclassical theory,  stochastic electrodynamics \cite{MS}). 
But the main  idea of all such alternatives was to treat the 
field in classical terms and to associate the observed discreteness of 
emission/absorbtion phenomena with the quantum nature of atoms and not with
the field itself. 

The approach we will discuss in this paper does not belong to this 
tradition, is much more radical and goes 
in the opposite direction. We will not 
try to make the field more classical. What we will try to do is to make 
it even {\it more quantum\/}  
by replacing classical parameters with eigenvalues. 

The field will be quantized at a one-particle level, but then extended 
to multi-particle systems. Obviously, it is not possible to include in a 
single paper all the possible tests of the new formalism one should perform. 
We will therefore concentrate on these points where
quantum electrodynamics produces results which are believed to be 
a consequence of the standard canonical quantization. 
Three areas should be checked first:

(i) Vacuum effects in atomic physics.

(ii) Emission of photons in entangled states. 

(iii) Boson statistics and the Planck law.

For the first two problems we shall choose the simplest approach, namely
first and second order perturbation theory. The Planck law will be discussed 
in a more detailed way. 
As we shall see the new theory is not completely equivalent to the
standard one, but the modifications one finds are surprisingly subtle
and in principle subject to experimental tests. 

In a separate paper we shall discuss  
perturbation theory to all orders, since then a kind of prescription for 
translating the old results into the new framework may appear. 

\section{Harmonic oscillator in superposition of frequencies}

We know that  frequency is typically associated with an
eigenvalue of some Hamiltonian or, which is basically the same, with
boundary conditions. A natural way of incorporating different
frequencies into a single harmonic oscillator is by means of the
{\it frequency operator\/} 
\be
\Omega=\sum_{\omega_k,j_k}\omega_k|\omega_k,j_k\rangle
\langle \omega_k,j_k|
\ee
where all $\omega_k\geq 0$. 
For simplicity we have limited the discussion to the discrete
spectrum but it is useful to include from the outset the possibility
of degeneracies, represented here by the additional discrete quantum
numbers $j_k$. The corresponding Hamiltonian is defined by
\be
H
&=&
\hbar\Omega\otimes \frac{1}{2}\big(a^{\dag}a+aa^{\dag}\big)
\ee
where $a=\sum_{n=0}^\infty\sqrt{n+1}|n\rangle\langle n+1|$. 
The eigenstates of $H$ are $|\omega_k,j_k,n\rangle$ and satisfy the required 
formula
\be
H|\omega_k,j_k,n\rangle= \hbar\omega_k\Big(n+\frac{1}{2}\Big)
|\omega_k,j_k,n\rangle
\ee
justifying our choice of $H$. 
The standard case of the oscillator whose frequency is just $\omega$ 
coresponds either to $\Omega=\omega\bbox 1$ or to the subspace
spanned by $|\omega_k,j_k,n\rangle$ with fixed $\omega_k=\omega$. 
Introducing the operators 
\be
a_{\omega_k,j_k}=|\omega_k,j_k\rangle\langle \omega_k,j_k|\otimes a
\ee
we find that 
\be 
H&=&
\frac{1}{2}\sum_{\omega_k,j_k}\hbar\omega_k
\Big(a_{\omega_k,j_k}^{\dag}a_{\omega_k,j_k} 
+
a_{\omega_k,j_k}a_{\omega_k,j_k}^{\dag}\Big).
\ee
The algebra of the oscillator is ``noncanonical":
\be
{[a_{\omega_k,j_k},a_{\omega_l,j_l}^{\dag}]}&=&
\delta_{\omega_k\omega_l}\delta_{j_kj_l}
|\omega_k,j_k\rangle\langle\omega_k,j_k|\otimes \bbox 1\\
a_{\omega_k,j_k}a_{\omega_l,j_l}&=&
\delta_{\omega_k\omega_l}\delta_{j_kj_l}(a_{\omega_k,j_k})^2\\
a_{\omega_k,j_k}^{\dag}a_{\omega_l,j_l}^{\dag}&=&
\delta_{\omega_k\omega_l}\delta_{j_kj_l}(a_{\omega_k,j_k}^{\dag})^2.
\ee
The dynamics in the Schr\"odinger picture is given by 
\be
i\hbar\partial_t|\Psi\rangle
&=&
H |\Psi\rangle=
\hbar\Omega\otimes \big(a^{\dag}a+\frac{1}{2}\bbox 1\big)
|\Psi\rangle.
\ee
In the Heisenberg picture we obtain the important formula (see
Appendix~\ref{A-imp})
\be
a_{\omega_k,j_k}(t)
&=&
e^{iHt/\hbar}a_{\omega_k,j_k}e^{-iHt/\hbar}\\
&=&
|\omega_k,j_k\rangle\langle \omega_k,j_k|
\otimes
e^{-i\omega_k t} a
=
e^{-i\omega_k t} a_{\omega_k,j_k}.
\label{exp}
\ee
Taking a general state
\be
|\psi\rangle=\sum_{\omega_k,j_k,n}\psi(\omega_k,j_k,n)|\omega_k,j_k\rangle
|n\rangle
\ee
we find that the average energy of the oscillator 
is 
\be
\langle H\rangle=
\langle\psi|H|\psi\rangle
=
\sum_{\omega_k,j_k,n}|\psi(\omega_k,j_k,n)|^2
\hbar\omega_k\Big(n+\frac{1}{2}\Big).
\ee
The average clearly looks as an average energy of an {\it ensemble of
different and independent oscillators\/}. The ground state of the
ensemble, i.e. the one with $\psi(\omega_k,j_k,n>0)=0$ 
has energy 
\be
\langle H\rangle=\frac{1}{2}\sum_{\omega_k,j_k}|\psi(\omega_k,j_k,0)|^2
\hbar\omega_k
\ee
which is finite if 
\be
\sum_{\omega_k,j_k}\psi(\omega_k,j_k,0)|\omega_k,j_k\rangle
\ee
belongs to the domain of $\Omega$.
The result is not surprising but still quite remarkable if one thinks
of the problem of field quantization. 

The very idea of quantizing the electromagnetic field, as put
forward by Born, Heisenberg, Jordan \cite{BHJ} and Dirac \cite{D},
is based on the observation that the mode decomposition of the
electromagnetic energy is analogous to the energy of an ensemble of
independent harmonic oscillators. In 1925, after the work of
Heisenberg, it was clear what to do: One had to replace each
classical oscillator by a quantum one. But since each oscillator had
a definite frequency, to have an infinite number of different
frequencies one needed an infinite number of oscillators. 
The price one payed for this assumption was the 
infinite energy of the electromagnetic vacuum. 

The infinity is regarded as an ``easy" one since one can get rid of
it by redefining the Hamiltonian and removing the infinite term. 
The result looks correct and many properties typical of a {\it
quantum\/} harmonic oscillator are indeed observed in electromagnetic
field. However, subtraction of infinite terms is in mathematics 
as forbidden as division by zero so to avoid evident absurdities one
is forced to invent various ad hoc regularizations whose only
justification is that otherwise the theory would not work.  
In larger perspective
(say, in cosmology) it is not at all clear that an infinite (or
arbitrarily cut off at the Planck scale) energy of
the vacuum does not lead to contradictions with observational data 
\cite{Lambda}. 
Finally, Dirac himself had never been fully satisfied by the theory
he created. 
As Weinberg put it, Dirac's  ``demand for a completely
finite theory is similar to a host of other aesthetic judgements that
theoretical physicists always need to make" \cite{Dreams}.

The oscillator that can exist in superpositions of different
frequencies is a natural candidate as a starting point for Dirac-type
field quantization. Symbolically, if the Heisenberg quantization is 
$p^2+\omega^2q^2\mapsto \hat p^2+\omega^2\hat q^2$, where $\omega$ is a
parameter, the new scheme is $p^2+\omega^2q^2\mapsto \hat p^2+\hat
\omega^2\hat q^2$, where $\hat \omega$ is an operator. Its spectrum
can be related to boundary conditions imposed on the fields. The
field now can exist in superposition of frequencies but the
superposition is meant in the quantum sense i.e. the field may
consist of (an indefinite
number) of oscillators with indefinite frequency. 
In this meaning the approach we propose is even ``more
quantum" than the standard one since $\omega$ is not a (classical)
parameter but an eigenvalue.

We do not need to remove the ground state energy since in the Hilbert
space of physical states the correction is finite. The question we
have to understand is whether one can obtain the well known quantum
properties of the radiation field by this type of quantization.

\section{Prelude: ``First quantization'' --- 
Field operators for free Maxwell fields}

The new quantization will be performed in two steps. In this section
we describe the first step, a kind of first quantization. In next
sections we shall perform an analogue of second quantization which
will lead to the final framework. It is essential that the
``second quantization" will not involve, in fact, 
any additional quantization but is simply a transition from one to many
oscillators.  

The energy and momentum operators of the field are defined in analogy to $H$
from the previous section
\be
H &=& 
\sum_{s,\kappa_\lambda}\hbar \omega_\lambda
|s,\vec \kappa_\lambda\rangle \langle s,\vec \kappa_\lambda|
\otimes \frac{1}{2}\Big(a^{\dag}a+a a^{\dag}\Big)\\
&=& 
\frac{1}{2}\sum_{s,\kappa_\lambda}\hbar \omega_\lambda
\Big(a_{s,\kappa_\lambda}^{\dag}a_{s,\kappa_\lambda} 
+a_{s,\kappa_\lambda} a_{s,\kappa_\lambda}^{\dag}\Big)\\
\vec P &=& 
\sum_{s,\kappa_\lambda}\hbar \vec \kappa_\lambda
|s,\vec \kappa_\lambda\rangle \langle s,\vec \kappa_\lambda|
\otimes \frac{1}{2}\Big(a^{\dag}a+a a^{\dag}\Big)\\
&=& 
\frac{1}{2}\sum_{s,\kappa_\lambda}\hbar \vec \kappa_\lambda
\Big(a_{s,\kappa_\lambda}^{\dag}a_{s,\kappa_\lambda} 
+a_{s,\kappa_\lambda} a_{s,\kappa_\lambda}^{\dag}\Big)
\ee
where $s=\pm 1$ corresponds to circular polarizations. Denote 
$P=(H/c,\vec P)$ and $P\cdot x=Ht-\vec P\cdot \vec x$.
We employ the standard Dirac-type definitions for mode quantization in
volume $V$
\be
\hat{\vec A}(t,\vec x)
&=&
\sum_{s,\kappa_\lambda}
\sqrt{\frac{\hbar}{2\omega_\lambda V}}
\Big(a_{s,\kappa_\lambda}
e^{-i\omega_\lambda t} \vec e_{s,\kappa_\lambda}
e^{i\vec \kappa_\lambda\cdot \vec x}
+
a^{\dag}_{s,\kappa_\lambda}e^{i\omega_\lambda t} 
\vec e^{\,*}_{s,\kappa_\lambda}
e^{-i\vec \kappa_\lambda\cdot \vec x}
\Big)\\
&=&
e^{iP\cdot x/\hbar} \hat{\vec A} e^{-iP\cdot x/\hbar}\\
\hat{\vec E}(t,\vec x)
&=&
i\sum_{s,\kappa_\lambda }
\sqrt{\frac{\hbar\omega_\lambda}{2V}}
\Big(
a_{s,\kappa_\lambda}e^{-i\omega_\lambda t} 
e^{i\vec \kappa_\lambda\cdot \vec x}
\vec e_{s,\kappa_\lambda}
-
a^{\dag}_{s,\kappa_\lambda}e^{i\omega_\lambda t} 
e^{-i\vec \kappa_\lambda\cdot \vec x}
\vec e^{\,*}_{s,\kappa_\lambda}
\Big)\\
&=&
e^{iP\cdot x/\hbar} \hat{\vec E} e^{-iP\cdot x/\hbar}\\
\\
\hat{\vec B}(t,\vec x)
&=&
i\sum_{s,\kappa_\lambda }
\sqrt{\frac{\hbar\omega_\lambda}{2V}}
\vec n_{\kappa_\lambda}
\times 
\Big(a_{s,\kappa_\lambda}e^{-i\omega_\lambda t} 
e^{i\vec \kappa_\lambda\cdot \vec x}
\vec e_{s,\kappa_\lambda}
-
a^{\dag}_{s,\kappa_\lambda}e^{i\omega_\lambda t} 
e^{-i\vec \kappa_\lambda\cdot \vec x}
\vec e^{\,*}_{s,\kappa_\lambda}
\Big)\\
&=&
e^{iP\cdot x/\hbar} \hat{\vec B} e^{-iP\cdot x/\hbar},
\ee
where
\be
a_{s,\vec \kappa_\lambda} &=&
|s,\vec \kappa_\lambda\rangle \langle s,\vec \kappa_\lambda|
\otimes a\label{a}\\
a^{\dag}_{s,\vec \kappa_\lambda} &=&
|s,\vec \kappa_\lambda\rangle \langle s,\vec \kappa_\lambda|
\otimes a^{\dag}.\label{a^dag}
\ee
For later purposes we introduce the notation
\be
[a_{s,\vec \kappa_\lambda},a_{s,\vec \kappa_\lambda}^{\dag}]
=
1_{s,\vec \kappa_\lambda}
=
|s,\vec \kappa_\lambda\rangle\langle s,\vec \kappa_\lambda|\otimes 
\bbox 1.\label{1_s}
\ee
Now take a state (say, in the Heisenberg picture)
\be
|\Psi\rangle
&=&
\sum_{s,\vec \kappa_\lambda,n}\Psi_{s,\vec \kappa_\lambda,n}
|s,\vec \kappa_\lambda,n\rangle\\
&=&
\sum_{s,\vec \kappa_\lambda}\Phi_{s,\vec \kappa_\lambda}
|s,\vec \kappa_\lambda\rangle|\alpha_{s,\vec \kappa_\lambda}\rangle
\ee
where $|\alpha_{s,\vec \kappa_\lambda}\rangle$ form a family of 
one-oscillator coherent states:
\be
a|\alpha_{s,\vec \kappa_\lambda}\rangle
=
\alpha_{s,\vec \kappa_\lambda}
|\alpha_{s,\vec \kappa_\lambda}\rangle
\ee
The averages of the field operators are 
\be
\langle\Psi|\hat{\vec A}(t,\vec x)|\Psi\rangle
&=&
\sum_{s,\kappa_\lambda }|\Phi_{s,\vec \kappa_\lambda}|^2
\sqrt{\frac{\hbar}{2\omega_\lambda V}}
\Big(
\alpha_{s,\kappa_\lambda}
e^{-i\kappa_\lambda\cdot x} 
\vec e_{s,\kappa_\lambda}
+
\alpha^*_{s,\kappa_\lambda}
e^{i\kappa_\lambda\cdot x}
\vec e^{\,*}_{s,\kappa_{\lambda}}
\Big)\\
\langle\Psi|\hat{\vec E}(t,\vec x)|\Psi\rangle
&=&
\sum_{s,\kappa_\lambda }|\Phi_{s,\vec \kappa_\lambda}|^2
\sqrt{\frac{\hbar\omega_\lambda}{2V}}
\Big(
\alpha_{s,\kappa_\lambda}
e^{-i\kappa_\lambda\cdot x}
\vec e_{s,\kappa_\lambda}
-
\alpha^*_{s,\kappa_\lambda}
e^{i\kappa_\lambda\cdot x}
\vec e^{\,*}_{s,\kappa_{\lambda}}
\Big)\\
\langle\Psi|\hat{\vec B}(t,\vec x)|\Psi\rangle
&=&
i\sum_{s,\kappa_\lambda }|\Phi_{s,\vec \kappa_\lambda}|^2
\sqrt{\frac{\hbar\omega_\lambda}{2V}}
\Big(
\alpha_{s,\kappa_\lambda}e^{-i\kappa_\lambda\cdot x}
\vec n_{\kappa_\lambda}
\times 
\vec e_{s,\kappa_\lambda}
-
\alpha^*_{s,\kappa_\lambda}e^{i\kappa_\lambda\cdot x}
\vec n_{\kappa_\lambda}
\times 
\vec e^{\,*}_{s,\kappa_{\lambda}}
\Big)
\ee
These are just the classical fields. More precisely, the fields look
like averages 
of monochromatic coherent states with probabilities 
$|\Phi_{s,\vec \kappa_\lambda}|^2$. The energy-momentum operators
satisfy also the standard relations (see Appendix~\ref{A-EP})
\be
H
&=&
\frac{1}{2}
\int_V d^3x
\Big(
\hat{\vec E}(t,\vec x)\cdot \hat{\vec E}(t,\vec x)
+
\hat{\vec B}(t,\vec x)\cdot \hat{\vec B}(t,\vec x)
\Big),\label{P1}\\
\vec P
&=&
\int_V d^3x \hat{\vec E}(t,\vec x)\times \hat{\vec B}(t,\vec x).
\label{P2}
\ee
It should be stressed, however, that these relations have a
completely different mathematical origin than in the usual formalism
where the integrals are necessary in order to make plane waves
into an orthonormal basis. Here orthogonality follows from the
presence of the projectors in the definition of
$a_{s,\kappa_\lambda}$ and the integration in itself is {\it
trivial\/} since
\be
\hat{\vec E}(t,\vec x)\cdot \hat{\vec E}(t,\vec x)
+
\hat{\vec B}(t,\vec x)\cdot \hat{\vec B}(t,\vec x)
&=&
\hat{\vec E}\cdot \hat{\vec E}
+
\hat{\vec B}\cdot \hat{\vec B}\\
\hat{\vec E}(t,\vec x)\times \hat{\vec B}(t,\vec x)
&=&
\hat{\vec E}\times \hat{\vec B}.
\ee
Therefore the role of the integral is simply to produce the factor
$V$ which cancels with $1/V$ arising from the term $1/\sqrt{V}$
occuring in the mode decomposition of the fields. 
To end this section let us note that 
\be
\langle \Psi|H|\Psi\rangle
&=&
\sum_{s,\kappa_\lambda }
\hbar\omega_\lambda 
|\Phi_{s,\kappa_\lambda}|^2
\Big(
|\alpha_{s,\kappa_\lambda}|^2
+\frac{1}{2}
\Big)\\
\langle \Psi|\vec P|\Psi\rangle
&=&
\sum_{s,\kappa_\lambda }
\hbar\vec \kappa_{\lambda}
|\Phi_{s,\kappa_\lambda}|^2
\Big(
|\alpha_{s,\kappa_\lambda}|^2
+\frac{1}{2}
\Big).
\ee
The contribution from the vacuum fluctuations is nonzero but {\it finite\/}. 
One can phrase the latter property also as follows. The noncanonical algebra 
of creation-annihilation operators satisfies the resolution of identity 
\be
\sum_{s,\kappa_\lambda }
[a_{s,\vec \kappa_\lambda},a_{s,\vec \kappa_\lambda}^{\dag}]
=
\bbox 1\label{r-id}
\ee
wheras the canonical algebra would impliy 
\be
\sum_{s,\kappa_\lambda }
[a_{s,\vec \kappa_\lambda},a_{s,\vec \kappa_\lambda}^{\dag}]
=
\infty \bbox 1.
\ee

\section{Spontaneous and stimulated emission: First version}

Some typically quantum optical phenomena occur already at the 
one-oscillator level. Below we shall see that spontaneous and stimulated 
emissions are a property of a single-oscillator description, although to 
have a more complete picture we need the multi-oscillator extension 
discussed in subsequent sections. 

Beginning with the dipole and rotating wave approximations (RWA)
we arrive at the Hamiltonian 
\be
H
&=&
\frac{1}{2}\hbar\omega_0\sigma_3
+
\frac{1}{2}\sum_{s,\vec \kappa_\lambda }\hbar\omega_\lambda 
\Big(
a_{s,\vec \kappa_\lambda}^{\dag}a_{s,\vec \kappa_\lambda} 
+a_{s,\vec \kappa_\lambda}a_{s,\vec \kappa_\lambda}^{\dag}\Big)
+
\hbar\omega_0d\sum_{s,\vec \kappa_\lambda }
\Big(
g_{s,\vec \kappa_\lambda}
a_{s,\vec \kappa_\lambda} \sigma_+
+
g_{s,\vec \kappa_\lambda}^*
a_{s,\vec \kappa_\lambda}^{\dag}\sigma_-
\Big)
\ee
where $d\vec u=\langle + |\hat{\vec d}|-\rangle$ is the matrix
element of the dipole
moment evaluated between the excited and ground states, 
and $g_{s,\vec \kappa_\lambda}=i\sqrt{\frac{1}{2\hbar\omega_\lambda V}} 
\vec e_{s,\vec \kappa_\lambda}\cdot \vec u$. The Hamiltonian represents a
two-level atom located at $\vec x_0=0$. 

The Hamiltonian in the interaction picture has the well known form
\be
H_I
&=&
\hbar\omega_0d\sum_{s,\vec \kappa_\lambda }
\Big(
g_{s,\vec \kappa_\lambda}e^{i(\omega_0-\omega_\lambda)t}
a_{s,\vec \kappa_\lambda} \sigma_+
+
g_{s,\vec \kappa_\lambda}^*e^{-i(\omega_0-\omega_\lambda)t}
a_{s,\vec \kappa_\lambda}^{\dag}\sigma_-
\Big).
\ee
Consider the initial state 
\be
|\Psi(0)\rangle
&=&
\sum_{s',\vec \kappa_{\lambda'},m}\Psi_{s',\vec \kappa_{\lambda'},m}
|s',\vec \kappa_{\lambda'},m,+\rangle\nonumber\\
&=&
\sum_{s',\vec \kappa'_{0}}
\Psi_{s',\vec \kappa'_{0},0}
|s',\vec \kappa'_{0},0,+\rangle
+
\sum_{s',\vec \kappa'_{n}}\Psi_{s',\vec \kappa'_{n},n}
|s',\vec \kappa'_{n},n,+\rangle.
\ee
The states corresponding to $n=0$ play a role of a
vacuum. As a 
consequence the vacuum is not represented here by a unique vector,
but rather by a subspace of the Hilbert space of states. 
Energy of the general vacuum state
\be
|\Psi\rangle
&=&
\sum_{s,\vec \kappa_{\lambda},\pm}\Psi_{s,\vec \kappa_{\lambda},0,\pm}
|s,\vec \kappa_{\lambda},0,\pm\rangle
\ee
is related to the density of modes 
$
\rho(\vec\kappa_{\lambda})=
\sum_{s,\pm}|\Psi_{s,\vec\kappa_{\lambda},0,\pm}|^2
$
and is, therefore, state dependent.

In order to estimate the probabilities of spontaneous and stimulated
emissions we can use the first-order time-dependent perturbation
theory \cite{Haken} and arrive at
\be
|\Psi(t)\rangle
&=&
|\Psi(0)\rangle
\nonumber\\
&\pp =&
+
\omega_0d
\sum_{s,\vec \kappa_0  }
\frac{e^{-i(\omega_0-\omega_{\lambda_0})t}-1}{\omega_0-\omega_\lambda}
\Psi_{s,\vec \kappa_{\lambda_0},0}
g_{s,\vec \kappa_{\lambda_0}}^*
|s,\vec \kappa_{\lambda_0},1,-\rangle\nonumber\\
&\pp =&
+
\omega_0d
\sum_{s,\vec \kappa_n  }
\frac{e^{-i(\omega_0-\omega_{\lambda_n})t}-1}{\omega_0-\omega_{\lambda_n}}
\Psi_{s,\vec \kappa_{n},n}
\sqrt{n+1}
g_{s,\vec \kappa_n}^*
|s,\vec \kappa_{n},n+1,-\rangle.
\ee
One recognizes here the well known contributions from spontaneous
and stimulated emissions. It should be stressed that although the
final result looks familiar, the mathematical details behind the calculation
are different from what we are accustomed to. For example, instead of
\be
a_{s_1,\vec \kappa_1}^{\dag}|s,\vec \kappa,m\rangle
\sim |s_1,\vec \kappa_1,1;s,\vec \kappa,m\rangle,
\ee
which would hold in the standard formalism for $\vec\kappa_1\neq
\vec\kappa$, we get simply
\be
a_{s_1,\vec \kappa_1}^{\dag}|s,\vec \kappa,m\rangle=0,
\ee
a consequence of 
$a_{s_1,\vec \kappa_1}^{\dag}a_{s,\vec \kappa}^{\dag}=0$. 

Let us now look more closely at spontaneous emission (we take $n=0$).
The state vector is 
\be
|\Psi(t)\rangle
&=&
|\Psi(0)\rangle
\nonumber\\
&\pp =&
+
\omega_0d
\sum_{s,\vec \kappa_0  }
\frac{e^{-i(\omega_0-\omega_{\lambda_0})t}-1}{\omega_0-\omega_\lambda}
G_{s,\vec \kappa_{\lambda_0}}^*
|s,\vec \kappa_{\lambda_0},1,-\rangle\label{``G''}
\ee
where we have introduced the ``effective coupling terms'' 
\be
G_{s,\vec \kappa_{\lambda_0}}^*
=
\Psi_{s,\vec \kappa_{\lambda_0},0}
g_{s,\vec \kappa_{\lambda_0}}^*.
\ee
As we can see the result is mathematically equivalent to the standard one
but with the coupling constants automatically regularized by the presence of
the vacuum amplitude in $G_{s,\vec \kappa_{\lambda_0}}^*$. 

We apply the standard argument but with $g$'s replaced by 
$G$'s and  obtain the probability of
emission per time unit
\be
P&=&
2\pi\omega_0^2 d^2\sum_{s,\vec \kappa}
|\Psi_{s,\vec \kappa,0}
g_{s,\vec \kappa}|^2
\delta(\omega_0-\omega_{\vec\kappa})\label{P}.
\ee
Assuming for simplicity 
that density of vacuum modes is isotropic and polarization
independent we can write it as a function of frequency only, i.e.
\be
|\Psi_{s,\vec \kappa,0}|^2=F(\omega_{\vec\kappa})
\ee
and finally
\be
P&=&
2\pi\omega_0^2 d^2
F(\omega_0)\sum_{s,\vec \kappa}
|g_{s,\vec \kappa}|^2
\delta(\omega_0-\omega_{\vec\kappa})=F(\omega_0)P_{\rm old}.\label{old}
\ee
Here $P_{\rm old}$ is the emission rate obtained in the standard
theory. The nontrivial structure of the vacuum influences the
lifetime of the atom.  We shall return to this and related questions
later but first have to extend the formalism in a way allowing to
consider entangled states of light. 

More reliable estimates of the lifetime require more detailed
calculations that we postpone to a forthcoming paper. One should also keep 
in mind the possibility of an unisotropic vacuum caused by more complicated
boundary conditions such as those occuring in measurements of the 
Casimir force.

\section{``Second quantization''}

The Hilbert space of states of the field we have constructed is
spanned by vectors $|s,\vec \kappa,n\rangle$. Still there is no 
doubt that both in reality (and the standard formalism) 
there exist multiparticle entangled states such
as those spanned by tensor products of the form
\be
|+,\vec \kappa_1,1\rangle
|-,\vec \kappa_2,1\rangle,
\ee
and the similar. It seems that there is no reason to limit our
discussion to a {\it single\/} Hilbert space of a {\it single\/}
oscillator. What we have done so far was a quantization of the
electromagnetic field at the level of a ``one-particle" Hilbert
space.
Similarly to quantization of other physical systems 
the next step is to consider many (noninteracting) particles. 

The procedure is essentially clear. Having the one-particle
energy-momentum operators $P_a$  (i.e. generators of 4-translations in the 
1-particle Hilbert space) we define in the standard way their extensions 
to the Fock-type space
\be
{\cal P}_a
&=&P_a\nonumber\\
&\pp =&
\oplus\big(P_a\otimes\bbox 1+\bbox 1\otimes P_a\big)\nonumber\\
&\pp =&
\oplus \big(P_a\otimes\bbox 1\otimes\bbox 1
+\bbox 1\otimes P_a\otimes\bbox 1
+\bbox 1\otimes \bbox 1\otimes P_a\big)\nonumber\\
&\pp =&
\oplus\dots.
\ee
The $x$-dependence of fields is introduced similarly to the
one-particle level
\be
\vec {\cal F}(t,\vec x)
&=&
e^{i{\cal P}\cdot x/\hbar}
\vec {\cal F}
e^{-i{\cal P}\cdot x/\hbar}
\ee
but the field itself has yet to be defined. Assume 
\be
\vec {\cal F}
&=&
c_1\vec F\nonumber\\
&\pp =&
\oplus c_2\big(\vec F\otimes\bbox 1+\bbox 1\otimes \vec F\big)\nonumber\\
&\pp =&
\oplus c_3\big(\vec F\otimes\bbox 1\otimes\bbox 1
+\bbox 1\otimes \vec F\otimes\bbox 1
+\bbox 1\otimes \bbox 1\otimes \vec F\big)\nonumber\\
&\pp =&
\oplus\dots
\ee
where $c_k$ are constants discussed below, 
and $\vec F$ is $\hat {\vec A}$, $\hat {\vec E}$, or $\hat {\vec B}$.
The multi-oscillator annihilation operator associated with such fields 
must be therefore of the form 
\be
\bbox a_{s,\vec \kappa}
&=&
c_1a_{s,\vec \kappa}\nonumber\\
&\pp =&
\oplus c_2\big(a_{s,\vec \kappa}\otimes\bbox 1+
\bbox 1\otimes a_{s,\vec \kappa}\big)\nonumber\\
&\pp =&
\oplus c_3\big(a_{s,\vec \kappa}\otimes\bbox 1\otimes\bbox 1
+\bbox 1\otimes a_{s,\vec \kappa}\otimes\bbox 1
+\bbox 1\otimes \bbox 1\otimes a_{s,\vec \kappa}\big)\nonumber\\
&\pp =&
\oplus\dots.
\ee

Having two 1-particle operators, say $X$ and $Y$, one can easily 
establish a relation between the 1-particle commutator $[X,Y]$ and the 
commutator of the extensions $\cal X$, $\cal Y$:
\be
{[{\cal X},{\cal Y}]}
&=&c_1^2[X,Y]\nonumber\\
&\pp =&
\oplus c_2^2\big([X,Y]\otimes\bbox 1+\bbox 1\otimes [X,Y]\big)\nonumber\\
&\pp =&
\oplus c_3^2\big([X,Y]\otimes\bbox 1\otimes\bbox 1
+\bbox 1\otimes [X,Y]\otimes\bbox 1
+\bbox 1\otimes \bbox 1\otimes [X,Y]\big)\nonumber\\
&\pp =&
\oplus\dots.
\ee
The annihilation operators so defined satisfy therefore the algebra
\be
{[\bbox a_{s,\vec \kappa},\bbox a_{s',\vec \kappa\,'}^{\dag}]}
&=&0 \quad {\rm for}\,  (s,\vec \kappa)\neq (s',\vec \kappa\,'),\\
{[\bbox a_{s,\vec \kappa},\bbox a_{s,\vec \kappa\,}^{\dag}]}
&=&\bbox 1_{s,\vec \kappa},\\
{[\bbox a_{s,\vec \kappa},\bbox a_{s',\vec \kappa\,'}]}
&=&0\\
{[\bbox a_{s,\vec \kappa}^{\dag},\bbox a_{s',\vec \kappa\,'}^{\dag}]}
&=&0
\ee 
where the operator $\bbox 1_{s,\vec \kappa}$ is defined by
\be
\bbox 1_{s,\vec \kappa}
&=&c_1^2 1_{s,\vec \kappa}\nonumber\\
&\pp =&
\oplus c_2^2\big(1_{s,\vec \kappa}
\otimes\bbox 1+\bbox 1\otimes 1_{s,\vec \kappa}\big)\nonumber\\
&\pp =&
\oplus c_3^2\big(1_{s,\vec \kappa}
\otimes\bbox 1\otimes\bbox 1
+\bbox 1\otimes 1_{s,\vec \kappa}
\otimes\bbox 1
+\bbox 1\otimes \bbox 1\otimes 1_{s,\vec \kappa}\big)\nonumber\\
&\pp =&
\oplus\dots,
\ee
and $1_{s,\vec \kappa}$ is a single-oscillator operator (\ref{1_s}). 

As opposed to the single-oscillator case 
\be
\bbox a_{s,\vec \kappa}\bbox a_{s',\vec \kappa\,'}\neq 
\delta_{ss'}\delta_{\vec \kappa,\vec \kappa'}
(\bbox a_{s,\vec \kappa})^2.
\ee
An important property of the 
1-oscillator description was the resolution of identity (\ref{r-id}). The 
requirement that the same be valid at the multi oscillator level leads to 
$c_n=1/\sqrt{n}$.
In such a case one finds that 
\be
\bbox 1_{s,\vec \kappa}^2\neq \bbox 1_{s,\vec \kappa}
\ee
but nevertheless 
\be
\sum_{s,\vec \kappa}
\bbox 1_{s,\vec \kappa}=\bbox 1.\label{r of i}
\ee
Below we shall give another justification of this particular choice of 
$c_n$. 

We can finally write 
\be
\vec{\cal A}(t,\vec x)
&=&
\sum_{s,\kappa_\lambda}
\sqrt{\frac{\hbar}{2\omega_\lambda V}}
\Big(\bbox a_{s,\kappa_\lambda}
e^{-i\omega_\lambda t} \vec e_{s,\kappa_\lambda}
e^{i\vec \kappa_\lambda\cdot \vec x}
+
\bbox a^{\dag}_{s,\kappa_\lambda}e^{i\omega_\lambda t} 
\vec e^{\,*}_{s,\kappa_\lambda}
e^{-i\vec \kappa_\lambda\cdot \vec x}
\Big)\\
&=&
e^{i{\cal P}\cdot x/\hbar} \vec{\cal A} e^{-i{\cal P}\cdot x/\hbar}\\
\vec{\cal E}(t,\vec x)
&=&
i\sum_{s,\kappa_\lambda }
\sqrt{\frac{\hbar\omega_\lambda}{2V}}
\Big(
\bbox a_{s,\kappa_\lambda}e^{-i\omega_\lambda t} 
e^{i\vec \kappa_\lambda\cdot \vec x}
\vec e_{s,\kappa_\lambda}
-
\bbox a^{\dag}_{s,\kappa_\lambda}e^{i\omega_\lambda t} 
e^{-i\vec \kappa_\lambda\cdot \vec x}
\vec e^{\,*}_{s,\kappa_\lambda}
\Big)\\
&=&
e^{i{\cal P}\cdot x/\hbar} \vec{\cal E} e^{-i{\cal P}\cdot x/\hbar}\\
\\
\vec{\cal B}(t,\vec x)
&=&
i\sum_{s,\kappa_\lambda }
\sqrt{\frac{\hbar\omega_\lambda}{2V}}
\vec n_{\kappa_\lambda}
\times 
\Big(\bbox a_{s,\kappa_\lambda}e^{-i\omega_\lambda t} 
e^{i\vec \kappa_\lambda\cdot \vec x}
\vec e_{s,\kappa_\lambda}
-
\bbox a^{\dag}_{s,\kappa_\lambda}e^{i\omega_\lambda t} 
e^{-i\vec \kappa_\lambda\cdot \vec x}
\vec e^{\,*}_{s,\kappa_\lambda}
\Big)\\
&=&
e^{i{\cal P}\cdot x/\hbar} \vec{\cal B} e^{-i{\cal P}\cdot x/\hbar}.
\ee
These operators form a basis of the modified version of
nonrelativistic quantum optics.

Let us return for the moment to the case of a general $c_n$. 
A straightforward calculation shows that
\be
{\bf H}&=&
\frac{1}{2}
\int_V
d^3x
\Big(
\vec{\cal E}(t,\vec x)
\cdot
\vec{\cal E}(t,\vec x)
+
\vec{\cal B}(t,\vec x)
\cdot
\vec{\cal B}(t,\vec x)
\Big)
=
\frac{1}{2}\sum_{s,\kappa_\lambda}\hbar \omega_\lambda
\Big(\bbox a_{s,\kappa_\lambda}^{\dag}\bbox a_{s,\kappa_\lambda} 
+\bbox a_{s,\kappa_\lambda} \bbox a_{s,\kappa_\lambda}^{\dag}\Big)\\
&=&
\frac{1}{2}\sum_{s,\kappa_\lambda}\hbar \omega_\lambda
\Bigg[
c_1^2
\{a_{s,\vec \kappa},a^{\dag}_{s,\vec \kappa}\}
\nonumber\\
&\pp =&
\oplus c_2^2\Big(\{a_{s,\vec \kappa},a^{\dag}_{s,\vec \kappa}\}
\otimes\bbox 1+
\bbox 1\otimes \{a_{s,\vec \kappa},a^{\dag}_{s,\vec \kappa}\}
+
2a^{\dag}_{s,\vec \kappa}\otimes a_{s,\vec \kappa}
+
2a_{s,\vec \kappa}\otimes a^{\dag}_{s,\vec \kappa}
\Big)\nonumber\\
&\pp =&
\oplus c_3^2
\Big(\{a_{s,\vec \kappa},a^{\dag}_{s,\vec \kappa}\}\otimes\bbox 1\otimes\bbox 1
+
2a_{s,\vec \kappa}\otimes a^{\dag}_{s,\vec \kappa}\otimes\bbox 1
+
2a_{s,\vec \kappa}\otimes\bbox 1\otimes  a^{\dag}_{s,\vec \kappa}\nonumber\\
&\pp =&\pp +
+
2 a^{\dag}_{s,\vec \kappa}\otimes a_{s,\vec \kappa}\otimes\bbox 1
+
\bbox 1\otimes \{a_{s,\vec \kappa},a^{\dag}_{s,\vec \kappa}\}\otimes\bbox 1
+
2\bbox 1\otimes a_{s,\vec \kappa}\otimes  a^{\dag}_{s,\vec \kappa}\nonumber\\
&\pp =&\pp +
+
2 a^{\dag}_{s,\vec \kappa}\otimes \bbox 1\otimes a_{s,\vec \kappa}
+
2\bbox 1\otimes  a^{\dag}_{s,\vec \kappa}\otimes a_{s,\vec \kappa}
+
\bbox 1\otimes \bbox 1\otimes \{a_{s,\vec \kappa},a^{\dag}_{s,\vec \kappa}\}
\Big)\nonumber\\
&\pp =&
\oplus\dots.
\Bigg]
\ee
Comparing this with the generator of time translations 
\be
{\cal H}=c{\cal P}_0
&=&
\frac{1}{2}\sum_{s,\kappa_\lambda}\hbar \omega_\lambda
\Bigg[
\{a_{s,\vec \kappa},a^{\dag}_{s,\vec \kappa}\}
\nonumber\\
&\pp =&
\oplus \Big(\{a_{s,\vec \kappa},a^{\dag}_{s,\vec \kappa}\}
\otimes\bbox 1+
\bbox 1\otimes \{a_{s,\vec \kappa},a^{\dag}_{s,\vec \kappa}\}
\Big)\nonumber\\
&\pp =&
\oplus 
\Big(\{a_{s,\vec \kappa},a^{\dag}_{s,\vec \kappa}\}\otimes\bbox 1\otimes\bbox 1
+
\bbox 1\otimes \{a_{s,\vec \kappa},a^{\dag}_{s,\vec \kappa}\}\otimes\bbox 1
+
\bbox 1\otimes \bbox 1\otimes \{a_{s,\vec \kappa},a^{\dag}_{s,\vec \kappa}\}
\Big)\nonumber\\
&\pp =&
\oplus\dots
\Bigg]
\ee
we can see that there is a relation between $\cal H$ and $\bf H$ but the latter
contains terms describing interactions between the oscillators. 
The contribution from these interactions vanishes on vacuum states. 
Below, when we introduce the notion of a generalized coherent state, we will
be able to relate averages of  $\cal H$ and $\bf H$. 
In a similar way one can introduce the ``Pointing operator''
\be
{\bf P} &=&
\int_V
d^3x\,
\vec{\cal E}(t,\vec x)
\times
\vec{\cal B}(t,\vec x)
=
\frac{1}{2}\sum_{s,\kappa_\lambda}\hbar \vec\kappa_\lambda
\Big(\bbox a_{s,\kappa_\lambda}^{\dag}\bbox a_{s,\kappa_\lambda} 
+\bbox a_{s,\kappa_\lambda} \bbox a_{s,\kappa_\lambda}^{\dag}\Big).
\ee
Its relation to the generator of 3-translations $\vec {\cal P}$ is similar 
to this between $\cal H$ and $\bf H$. 

In the above construction the only element which is beyond a simple 
transition to many oscillators is the choice of $c_n$. For different 
choices of these constants we obtain different algebras of noncanonical
commutation relations and therefore also different quantization schemes. 
Several different ways of reasoning lead to $c_n=1/\sqrt{n}$ as we shall 
also see in the next sections. 

\section{Some particular states}

We assume that all the multi-oscillator states are symmetric with
respect to permutations of the oscillators. 

\subsection{Generalized coherent states}

For general $c_n$ an eigenstate of $\bbox a_{s,\kappa_\lambda}$ 
corresponding to the eigenvalue $\alpha_{s,\vec \kappa}$
is of the form
\be
|\bbox \alpha_{s,\vec \kappa}\rangle
&=&
f_1(s,\vec \kappa)|s,\vec \kappa,\alpha_{s,\vec \kappa}/c_1\rangle\nonumber\\
&\pp =&
\oplus
f_2(s,\vec \kappa)
|s,\vec \kappa,\alpha_{s,\vec \kappa}/(2c_2)\rangle
|s,\vec \kappa,\alpha_{s,\vec \kappa}/(2c_2)\rangle\nonumber\\
&\pp =&
\oplus
f_3(s,\vec \kappa)
|s,\vec \kappa,\alpha_{s,\vec \kappa}/(3c_3)\rangle
|s,\vec \kappa,\alpha_{s,\vec \kappa}/(3c_3)\rangle
|s,\vec \kappa,\alpha_{s,\vec \kappa}/(3c_3)\rangle\nonumber\\
&\pp =&
\oplus\dots
\ee
where 
\be
|s,\vec \kappa,\alpha_{s,\vec \kappa}\rangle=|s,\vec \kappa\rangle
|\alpha_{s,\vec \kappa}\rangle,
\ee
$\sum_k|f_k(s,\vec \kappa)|^2=1$,
and 
$a |\alpha_{s,\vec \kappa}\rangle=\alpha_{s,\vec \kappa}
|\alpha_{s,\vec \kappa}\rangle$. 
What is interesting not all $f_k$ have to be nonvanishing. 

The average ``energies'' of the field in the above eigenstate are
\be
\langle\bbox \alpha_{s,\vec \kappa}|
{\cal H}
|\bbox \alpha_{s,\vec \kappa}\rangle
&=&
\hbar\omega_{\vec \kappa}
|\alpha_{s,\vec \kappa}|^2
\sum_{k=1}^\infty\frac{1}{kc_k^2}|f_k(s,\vec \kappa)|^2
+
\frac{1}{2}\hbar\omega_{\vec \kappa}\sum_{k=1}^\infty k|f_k(s,\vec \kappa)|^2
\ee
and 
\be
\langle\bbox \alpha_{s,\vec \kappa}|
{\bf H}
|\bbox \alpha_{s,\vec \kappa}\rangle
&=&
\hbar\omega_{\vec \kappa}
|\alpha_{s,\vec \kappa}|^2
+
\frac{1}{2}\hbar\omega_{\vec \kappa}
\sum_{k=1}^\infty kc_k^2|f_k(s,\vec \kappa)|^2.
\ee
The two averages will differ only by the value of the vacuum contribution 
if $c_k=1/\sqrt{k}$ which leads back to the above mentioned choice of $c_k$. 
With this choice and taking the general combination of coherent states 
\be
|\bbox \Psi\rangle
=\sum_{s,\vec \kappa}\Phi_{s,\vec \kappa}|\bbox \alpha_{s,\vec \kappa}\rangle
\ee
we find 
\be
\langle\bbox \Psi|
{\cal H}
|\bbox \Psi\rangle
&=&
\sum_{s,\vec \kappa}
\hbar\omega_{\vec \kappa}
|\Phi_{s,\vec \kappa}|^2
|\alpha_{s,\vec \kappa}|^2
+
\frac{1}{2}
\sum_{s,\vec \kappa}
\hbar\omega_{\vec \kappa}
|\Phi_{s,\vec \kappa}|^2
\sum_{k=1}^\infty k|f_k(s,\vec \kappa)|^2
\ee
and 
\be
\langle \bbox \Psi|
{\bf H}
|\bbox \Psi\rangle
&=&
\sum_{s,\vec \kappa}\hbar\omega_{\vec \kappa}
|\Phi_{s,\vec \kappa}|^2|\alpha_{s,\vec \kappa}|^2
+
\frac{1}{2}\sum_{s,\vec \kappa}
\hbar\omega_{\vec \kappa}|\Phi_{s,\vec \kappa}|^2.
\ee

\subsection{Vacuum}

Similarly to the one-oscillator case the traditional notion of a
vacuum state is replaced in our formalism by a vacuum {\it
subspace\/} consisting of all the vectors of the form
\be
|\bbox \Psi\rangle
&=&
\sum_{s,\vec \kappa_{\lambda}}\Psi^{(1)}_{s,\vec \kappa_{\lambda},0}
|s,\vec \kappa_{\lambda},0\rangle\nonumber\\
&\pp =&
\oplus
\sum_{s_j,\vec \kappa_{\lambda_j}}\Psi^{(2)}_{s_1,s_2,\vec
\kappa_{\lambda_1},\vec\kappa_{\lambda_2},0,0}
|s_1,\vec \kappa_{\lambda_1},0\rangle
|s_2,\vec \kappa_{\lambda_2},0\rangle\nonumber\\
&\pp =&
\oplus
\sum_{s_j,\vec \kappa_{\lambda_j}}\Psi^{(3)}_{s_1,s_2,s_3\vec
\kappa_{\lambda_1},\vec\kappa_{\lambda_2},\vec\kappa_{\lambda_3},0,0,0}
|s_1,\vec \kappa_{\lambda_1},0\rangle
|s_2,\vec \kappa_{\lambda_2},0\rangle
|s_3,\vec \kappa_{\lambda_3},0\rangle\nonumber\\
&\pp =&
\oplus\dots
\ee
It seems that there is no reason for introducing the standard ``vacuum state"
understood as the cyclic vector of the GNS construction. 

In the discussion of various vacuum phenomena (e.g. spontaneous
emission) we will assume for simplicity 
that all the oscillators are ``embedded" in
identical vacua i.e. the multi-oscillator vacuum is of the form 
\be
|\bbox \Psi\rangle
&=&
\sqrt{p_1}|\phi\rangle\nonumber\\
&\pp =&
\oplus
\sqrt{p_2}|\phi\rangle|\phi\rangle
\nonumber\\
&\pp =&
\oplus
\sqrt{p_3}
|\phi\rangle|\phi\rangle|\phi\rangle\nonumber\\
&\pp =&
\oplus\dots\label{multi-vac}
\ee
The average energy of the free-field vacuum state is therefore
\be
\overline{{\cal H}}=
\langle \bbox \Psi|
{\cal H}
|\bbox \Psi\rangle
=
\sum_{n=1}^\infty
np_n\langle\phi|
H
|\phi\rangle=\overline{n} \overline{H} 
\ee
where $\overline{n}$ and $\overline{H}$ are, respectively, the
average number of oscillators and the average energy of a single
oscillator. Again no problem with infinite vacuum energy is found. 
Obviously, one can contemplate also other vacua, say, in
entangled or mixed states.

\subsection{Multi-oscillator vs multi-photon states}

The coherent states we have introduced at the one-oscillator level
involve superpositions of different excited states. We know that in
the traditional approach the transition between two such states, say,
\be
|s,\vec\kappa,2\rangle
\to 
|s,\vec\kappa,0\rangle
\ee
is interpreted as an 
absorbtion (by some system) of two photons. 
In the new formulation the problem is more
complicated since the ``2-photon" absorbtion may be represented also
by 
\be
|s,\vec\kappa,1\rangle|s,\vec\kappa,1\rangle
\to
|s,\vec\kappa,0\rangle|s,\vec\kappa,0\rangle.
\ee
The two types of transitions do not represent the same process and
the two final states are physically distinguishable since their
energies are different. 
Indeed, 
\be
{\cal H}|s,\vec\kappa,0\rangle
&=&
H|s,\vec\kappa,0\rangle=
\frac{1}{2}\hbar \omega_{\vec\kappa}
|s,\vec\kappa,0\rangle
\ee
whereas
\be
{\cal H}
|s,\vec\kappa,0\rangle|s,\vec\kappa,0\rangle
&=&
(H\otimes \bbox 1+\bbox 1\otimes H)
|s,\vec\kappa,0\rangle|s,\vec\kappa,0\rangle
\nonumber\\
&=&
\hbar \omega_{\vec\kappa}
|s,\vec\kappa,0\rangle|s,\vec\kappa,0\rangle.
\ee
The notion of a 2-photon state becomes therefore somewhat ambiguous. 
To make it more precise one has to formulate a photodetection
theory within the new framework. In what follows we shall try to avoid the
use of the word ``photon" and will talk about ``light quanta" and 
``multi-oscillator" (or $n$-oscillator) and ``higher-excited" (or
$n$-th excited) states of light. 

The 2-oscillator states 
\be
|\bbox \Psi_\pm\rangle
&=&
\sum_{\vec \kappa_{1},\vec \kappa_{2}}
\Psi^{(2)}_{\vec\kappa_{1},\vec\kappa_{2},n}
\Big(
|+,\vec\kappa_1,n\rangle|-,\vec\kappa_2,n\rangle
\pm
|-,\vec\kappa_1,n\rangle|+,\vec\kappa_2,n\rangle
\Big),
\ee
satisfying 
\be
\Psi^{(2)}_{\vec\kappa_{1},\vec\kappa_{2},n}
=
\pm
\Psi^{(2)}_{\vec\kappa_{2},\vec\kappa_{1},n}
\ee
are (for any $n>0$) perfectly justified generalizations of the
standard 2-photon maximally entangled state. We shall later see that
although such ``higher excited photons" (i.e. $n>1$) 
are in principle possible, they are not produced in a two-photon
spontaneous emission (at least up to second-order perturbative
effects). The technical reason for this is the same as in the
ordinary formalism and is related to properties of the annihilation
operator $a$. 

\section{Spontaneous emission of a ``single photon"}

In this section we shall again consider the spontaneous emission of
light within the two-level-atom approximation. The example illustrates
some pecularities of the multi-oscillator formulation.

Denote by ${\cal H}_F$ the multi-oscillator Hamiltonian of the free
field we have discussed in the previous two sections. The dipole and
RWA Hamiltonian of the 2-level atom interacting with quantized
electromagnetic field is now
\be
{\cal H}
&=&
\frac{1}{2}\hbar\omega_0\sigma_3
+
{\cal H}_F
+
\hbar\omega_0d\sum_{s,\vec \kappa_\lambda }
\Big(
g_{s,\vec \kappa_\lambda}
\bbox a_{s,\vec \kappa_\lambda} \sigma_+
+
g_{s,\vec \kappa_\lambda}^*
\bbox a_{s,\vec \kappa_\lambda}^{\dag}\sigma_-
\Big).
\ee
Similarly to the one-oscillator case one has
\be
e^{i{\cal H}_Ft/\hbar}\bbox a_{s,\vec \kappa}e^{-i{\cal H}_Ft/\hbar}
=
e^{-i\omega_{\vec \kappa} t} \bbox a_{s,\vec \kappa}
\ee
and therefore the interaction-picture Hamiltonian 
is 
\be
{\cal H}_I
&=&
\hbar\omega_0d\sum_{s,\vec \kappa_\lambda }
\Big(
g_{s,\vec \kappa_\lambda}e^{i(\omega_0-\omega_\lambda)t}
\bbox a_{s,\vec \kappa_\lambda} \sigma_+
+
g_{s,\vec \kappa_\lambda}^*e^{-i(\omega_0-\omega_\lambda)t}
\bbox a_{s,\vec \kappa_\lambda}^{\dag}\sigma_-
\Big).
\ee
The first pecularity we encounter is the fact that the Hamiltonian is
{\it block diagonal\/} with respect to $\oplus$ and therefore does
not have nonvanishing matrix elements between spaces corresponding to
different numbers of oscillators. As a result the interaction cannot
change the number of oscillators, a property of crucial importance
for statistical properties of light as we shall see in the context of
the Planck blackbody radiation law. 

Consider the initial state
\be
|\bbox \Psi(0)\rangle
&=&
\sqrt{p_1}|+\rangle\sum_{s,\vec \kappa}\phi_{s,\vec \kappa}
|s,\vec \kappa,0\rangle\nonumber\\
&\pp =&
\oplus
\sqrt{p_2}|+\rangle
\sum_{s_1,\vec \kappa_1}
\sum_{s_2,\vec \kappa_2}
\phi_{s_1,\vec \kappa_1}\phi_{s_2,\vec \kappa_2}
|s_1,\vec \kappa_1,0\rangle
|s_2,\vec \kappa_2,0\rangle\nonumber\\
&\pp =&
\oplus
\sqrt{p_3}|+\rangle\sum_{s_1,\vec \kappa_1}
\sum_{s_2,\vec \kappa_2}
\sum_{s_3,\vec \kappa_3}
\phi_{s_1,\vec \kappa_1}
\phi_{s_2,\vec \kappa_2}
\phi_{s_3,\vec \kappa_3}
|s_1,\vec \kappa_1,0\rangle
|s_2,\vec \kappa_2,0\rangle
|s_3,\vec \kappa_3,0\rangle
\ee
The first-order perturbation theory yields (it is instructive to 
keep again the constants $c_n$ arbitrary)
\be
|\Psi(t)\rangle
&=&
|\Psi(0)\rangle
+
\omega_0d c_1\sqrt{p_1}
\sum_{s,\vec \kappa}
\frac{e^{-i(\omega_0-\omega_{\vec \kappa})t}-1}
{\omega_0-\omega_{\vec \kappa}}
g_{s,\vec \kappa}^*
\phi_{s,\vec \kappa}
|-\rangle|s,\vec \kappa,1\rangle\nonumber\\
&\pp =&
\oplus
\omega_0d
c_2\sqrt{p_2}
|-\rangle
\Bigg(
\sum_{s_1,\vec \kappa_1}
\frac{e^{-i(\omega_0-\omega_{\vec \kappa_1})t}-1}
{\omega_0-\omega_{\vec \kappa_1}}
g_{s_1,\vec \kappa_1}^*
\phi_{s_1,\vec \kappa_1}
|s_1,\vec \kappa_{1},1\rangle
|\phi\rangle
+
\sum_{s_2,\vec \kappa_2}
\frac{e^{-i(\omega_0-\omega_{\vec \kappa_2})t}-1}
{\omega_0-\omega_{\vec \kappa_2}}
g_{s_2,\vec \kappa_2}^*
\phi_{s_2,\vec \kappa_2}|\phi\rangle
|s_2,\vec \kappa_{2},1\rangle\Bigg)\nonumber\\
&\pp =&
\oplus\dots
\ee
As we can see, the ``single-photon" emission can be realized in an
infinite number of different ways. In the 1-oscillator subspace the
oscillator simply gets excited to the 1-st excited state. The
probability amplitude for this process is proportional to the
probability amplitude that the field is found in a 1-oscillator
state. In the 2-oscillator subspace there are two possibilities:
Either the first oscillator gets excited and the second one remains
in the ground state, or the other way around. The probability amplitude for
this process is proportional to to the probability amplitude that the
field is found in a 2-oscillator state. And so on.

Repeating the argument given for a single-oscillator description,
assuming the isotropy and polarization-independence of the
vacuum mode density, 
we arrive at the spontaneous emission rate of the form 
\be
P=F(\omega_0)P_{\rm old}\sum_{n=1}^\infty nc_n^2p_n
\ee
where $F(\omega_{\vec \kappa})=|\phi_{s,\vec \kappa}|^2$. As we can see the 
choice $c_n=1/\sqrt{n}$ plays again a special role since then  
\be
P=F(\omega_0)P_{\rm old},
\ee
that is, the result is the same as in the single-oscillator description. 

\section{Spontaneous emission of ``two photons"}

In what follows we will start with the Hamiltonian
$H=H_0+V$, where 
\be
H_0 &=& 
H_A
+
{\cal H}_F\\
V &=&
-\frac{e}{m}
\vec{\cal A}(\vec x)\cdot \vec p
\nonumber\\
&=&
-\frac{e}{m}
\sum_{s,\vec\kappa}
\sqrt{\frac{\hbar}{2\omega_{\vec\kappa} V}}
\Big(\bbox a_{s,\vec\kappa}
e^{i\vec \kappa\cdot \vec x}
\vec e_{s,\vec\kappa}\cdot\vec p
+
\bbox a^{\dag}_{s,\vec\kappa}
e^{-i\vec \kappa\cdot \vec x}
\vec e^{\,*}_{s,\vec\kappa}\cdot\vec p
\Big).
\ee
$H_A$ is the full (i.e. infinite-level) 
Hamiltonian describing an atom and ${\cal H}_F$ is the 
free-field Hamiltonian dicussed in Sec.~V and obtained by the 
multi-oscillator extension of the one-oscillator Hamiltonian introduced in
Sec.~III. To simplify notation we shall denote the sum and the integral over, 
respectively, the discrete and the continuous parts of the spectrum of  
$H_A$ by the sum $\sum_c$. We are not making the rotating wave approximation. 
In the dipole approximation we set $\vec x=0$. We shall also keep the 
constants $c_n$ arbitrary. 

\subsection{Two different light-quanta in 2-oscillator subspace}

In this subsection we will use the second-order perturbation 
theory to compute the amplitude 
\be
\langle b|\langle s_1,\vec \kappa_1,1|
\langle s_2,\vec \kappa_2,1|U(t_f,t_i)|a\rangle|\Psi\rangle
\ee
where the states $|s_k,\vec \kappa_k,1\rangle$, $k=1,2$, are
orthogonal, $U(t_f,t_i)$ is the evolution operator mapping the
initial state at time $t_i$ into the final state at time $t_f$, 
$|\bbox \Psi\rangle$ is a vacuum state (\ref{multi-vac}), and 
$|a\rangle$, $|b\rangle$ are two bound states of the atomic
Hamiltonian $H_A$. 

The fact that the interaction term does not change the number of
oscillators reduces the above amplitude to its 2-oscillator 
counterpart
\be
\sqrt{p_2}\langle b|\langle s_1,\vec \kappa_1,1|
\langle s_2,\vec \kappa_2,1|U(t_f,t_i)|a\rangle|\phi\rangle
|\phi\rangle
\ee
Using the standard perturbative techniques we obtain the second-order
approximation \cite{C-T} (see Appendix~\ref{A-2in2})
\be
{}&{}&
\langle b|\langle s_1,\vec \kappa_1,1|
\langle s_2,\vec \kappa_2,1|U^{(2)}(t_f,t_i)|a\rangle
|\phi\rangle|\phi\rangle\nonumber\\
&{}&
=
-c_2^2\frac{2\pi ie^2}{m^2}
\sqrt{\frac{\hbar}{2\omega_{\vec \kappa_2} V}}
\sqrt{\frac{\hbar}{2\omega_{\vec \kappa_1} V}}
\phi_{s_1,\vec \kappa_1}\phi_{s_2,\vec \kappa_2}
\sum_{c}\frac{
\big(
\vec e^{\,*}_{s_2,\vec \kappa_2}\cdot\vec p_{bc}
\big)
\big(
\vec e^{\,*}_{s_1,\vec \kappa_1}\cdot\vec p_{ca}
\big)}{E_{a,\vec \kappa_1,0,\vec \kappa_2,0}
-E_{c,\vec \kappa_1,1,\vec \kappa_2,0}+i0_+}
\delta^{(T)}(E_{a,\vec \kappa_1,\vec \kappa_2}-
E_{b,\vec \kappa_1,1,\vec \kappa_2,1})
\nonumber\\
&{}&\pp =
-c_2^2\frac{2\pi ie^2}{m^2}
\sqrt{\frac{\hbar}{2\omega_{\vec \kappa_2} V}}
\sqrt{\frac{\hbar}{2\omega_{\vec \kappa_1} V}}
\phi_{s_1,\vec \kappa_1}\phi_{s_2,\vec \kappa_2}
\sum_{c}\frac{
\big(
\vec e^{\,*}_{s_1,\vec \kappa_1}\cdot\vec p_{bc}
\big)
\big(
\vec e^{\,*}_{s_2,\vec \kappa_2}\cdot\vec p_{ca}
\big)}{E_{a,\vec \kappa_1,0,\vec \kappa_2,0}
-E_{c,\vec \kappa_1,0,\vec \kappa_2,1}+i0_+}
\delta^{(T)}(E_{a,\vec \kappa_1,\vec \kappa_2}-
E_{b,\vec \kappa_1,1,\vec \kappa_2,1})
\ee
where
$\delta^{(T)}(E-E')=[\pi(E-E')]^{-1}\sin\big((E-E')T/2\hbar\big)$, 
$\vec p_{bc}=\langle b|\vec p|c\rangle$, and 
$\vec p_{ca}=\langle c|\vec p|a\rangle$. 

The energies occuring in the above expression are (ground-state
energies are {\it not\/} removed!)
\be
E_{a,\vec \kappa_1,0,\vec \kappa_2,0}
&=&
E_a+\frac{1}{2}\hbar\omega_{\vec \kappa_1}+
\frac{1}{2}\hbar\omega_{\vec \kappa_2}\\
E_{b,\vec \kappa_1,1,\vec \kappa_2,1}
&=&
E_b+\big(1+\frac{1}{2}\big)\hbar\omega_{\vec \kappa_1}+
\big(1+\frac{1}{2}\big)\hbar\omega_{\vec \kappa_2}\\
E_{c,\vec \kappa_1,1,\vec \kappa_2,0}
&=&
E_c+\big(1+\frac{1}{2}\big)\hbar\omega_{\vec \kappa_1}+
\frac{1}{2}\hbar\omega_{\vec \kappa_2}\\
E_{c,\vec \kappa_1,0,\vec \kappa_2,1}
&=&
E_c+\frac{1}{2}\hbar\omega_{\vec \kappa_1}+
\big(1+\frac{1}{2}\big)\hbar\omega_{\vec \kappa_2}
\ee
The net result is the following
\be
{}&{}&\langle b|\langle s_1,\vec \kappa_1,1|
\langle s_2,\vec \kappa_2,1|U(t_f,t_i)|a\rangle|\Psi\rangle
\nonumber\\
&{}&
\approx
-c_2^2\sqrt{p_2}
\frac{2\pi ie^2}{m^2}
\sqrt{\frac{\hbar}{2\omega_{\vec \kappa_2} V}}
\sqrt{\frac{\hbar}{2\omega_{\vec \kappa_1} V}}
\phi_{s_1,\vec \kappa_1}\phi_{s_2,\vec \kappa_2}
\sum_{c}\frac{
\big(
\vec e^{\,*}_{s_2,\vec \kappa_2}\cdot\vec p_{bc}
\big)
\big(
\vec e^{\,*}_{s_1,\vec \kappa_1}\cdot\vec p_{ca}
\big)}
{E_a-E_c-\hbar\omega_{\vec \kappa_1}+i0_+}
\delta^{(T)}(E_a-E_b-\hbar\omega_{\vec \kappa_1}
-\hbar\omega_{\vec \kappa_2})
\nonumber\\
&{}&\pp =
-c_2^2\sqrt{p_2}\frac{2\pi ie^2}{m^2}
\sqrt{\frac{\hbar}{2\omega_{\vec \kappa_2} V}}
\sqrt{\frac{\hbar}{2\omega_{\vec \kappa_1} V}}
\phi_{s_1,\vec \kappa_1}\phi_{s_2,\vec \kappa_2}
\sum_{c}\frac{
\big(
\vec e^{\,*}_{s_1,\vec \kappa_1}\cdot\vec p_{bc}
\big)
\big(
\vec e^{\,*}_{s_2,\vec \kappa_2}\cdot\vec p_{ca}
\big)}{E_a-E_c-\hbar\omega_{\vec \kappa_2}+i0_+}
\delta^{(T)}(E_a-E_b-\hbar\omega_{\vec \kappa_1}
-\hbar\omega_{\vec \kappa_2})
\nonumber
\ee
Let us note that the amplitude is symmetric with respect to
permutation of states of the two oscillators:
\be
&\langle b|\langle s_1,\vec \kappa_1,1|
\langle s_2,\vec \kappa_2,1|U(t_f,t_i)|a\rangle|\Psi\rangle
=\langle b|\langle s_2,\vec \kappa_2,1|
\langle s_1,\vec \kappa_1,1|U(t_f,t_i)|a\rangle|\Psi\rangle
\nonumber
\ee
\subsection{Two different light-quanta in 3-oscillator subspace}

Consider the amplitudes
\be
{}&{}&
\langle b|\langle s_1,\vec \kappa_1,1|
\langle s_2,\vec \kappa_2,1|\langle\phi|U(t_f,t_i)|a\rangle|\Psi\rangle\\
{}&{}&
\langle b|\langle s_1,\vec \kappa_1,1|\langle\phi|
\langle s_2,\vec \kappa_2,1|U(t_f,t_i)|a\rangle|\Psi\rangle\\
{}&{}&
\langle b|\langle\phi|\langle s_1,\vec \kappa_1,1|
\langle s_2,\vec \kappa_2,1|U(t_f,t_i)|a\rangle|\Psi\rangle
\ee
In the framework we propose it is necessary to include the
contributions of this type arising from all the possible numbers of
oscillators. 

It is again sufficient to restrict the analysis to 
\be
\sqrt{p_3}\langle b|\langle s_1,\vec \kappa_1,1|
\langle s_2,\vec \kappa_2,1|\langle\phi|U(t_f,t_i)|a\rangle|\phi\rangle
|\phi\rangle|\phi\rangle
\ee
and similarly with the other two amplitudes. 
In second-order perturbation theory (see Appendix~\ref{A-2in3})
\be
c_3^{-2}\langle b|\langle s_1,\vec \kappa_1,1|
\langle s_2,\vec \kappa_2,1|\langle\phi|U^{(2)}(t_f,t_i)|a\rangle|\phi\rangle
|\phi\rangle|\phi\rangle
=
c_2^{-2}\langle b|\langle s_1,\vec \kappa_1,1|
\langle s_2,\vec \kappa_2,1|U^{(2)}(t_f,t_i)|a\rangle|\phi\rangle
|\phi\rangle.
\ee
The result is therefore essentially identical to the one obtained for the
2-oscillator subspace. 

A closer look at the derivation of the 3-oscillator contribution shows
that (i) exactly the same will happen for any number of oscillators
and
(ii) the second-order amplitude describes an emission of at most two 
quanta.

\subsection{Comparison with the standard formalism}

It is instructive to compare the result we have obtained with the
second-order calculation performed by means of the ordinary quantum optics
formalism. Let 
\be
|s_1,\vec \kappa_1;s_2,\vec \kappa_2\rangle
&=&
|s_2,\vec \kappa_2;s_1,\vec \kappa_1\rangle
=
\bbox a^{\dag}_{s_1,\vec\kappa_1}\bbox a^{\dag}_{s_2,\vec\kappa_2}
|0\rangle
\ee
be the two-photon state of the standard formalism, $\bbox
a^{\dag}_{s,\vec\kappa}$ the standard creation operator and
$|0\rangle$ the vacuum state. Then
\be
{}&{}&
\langle b|\langle s_1,\vec \kappa_1;
s_2,\vec \kappa_2|U^{(2)}(t_f,t_i)|a\rangle
|0\rangle
\nonumber\\
&{}&=
-\frac{2\pi ie^2}{m^2}
\sum_{c}
\sqrt{\frac{\hbar}{2\omega_{\vec\kappa_1} V}}
\sqrt{\frac{\hbar}{2\omega_{\vec\kappa_2} V}}
\big(\vec e^{\,*}_{s_1,\vec\kappa_1}\cdot\vec p_{bc}\big)
\big(\vec e^{\,*}_{s_2,\vec\kappa_2}\cdot\vec p_{ca}\big)
\frac{\delta^{(T)}(E_a-E_b-\hbar\omega_{\kappa_1}-\hbar\omega_{\kappa_2})}
{E_a-E_c-\hbar\omega_{\kappa_2}+i0_+}
\nonumber\\
{}&{}&\pp =
-\frac{2\pi ie^2}{m^2}
\sum_{c}
\sqrt{\frac{\hbar}{2\omega_{\vec\kappa_2} V}}
\sqrt{\frac{\hbar}{2\omega_{\vec\kappa_1} V}}
\big(\vec e^{\,*}_{s_2,\vec\kappa_2}\cdot\vec p_{bc}\big)
\big(\vec e^{\,*}_{s_1,\vec\kappa_1}\cdot\vec p_{ca}\big)
\frac{\delta^{(T)}(E_a-E_b-\hbar\omega_{\kappa_1}-\hbar\omega_{\kappa_2})}
{E_a-E_c-\hbar\omega_{\kappa_1}+i0_+}.
\nonumber
\ee
This is precisely the same expression that occurs in the modified 
amplitudes. 

\subsection{Probability of spontaneous emission of two quanta}

In the subspace corresponding to $n$ oscillators the ``two-photon"
emission can take place in 
\be
\left(\begin{array}{c}
n\\
2
\end{array}
\right)=\frac{n(n-1)}{2}
\ee
different ways. 
Taking into account probability amplitudes associated with all the
$n$-oscillator subspaces, $n>1$, and the fact that the two quanta 
can be emitted in two different orders, we obtain 
\be
p(s_1,\vec\kappa_1,s_2,\vec\kappa_2)
&=&
2\sum_{n=2}^\infty \frac{n(n-1)}{2}c_n^4p_n
|\phi_{s_1,\vec \kappa_1}|^2
|\phi_{s_2,\vec \kappa_2}|^2
p(s_1,\vec\kappa_1,s_2,\vec\kappa_2)_{\rm old}
\ee
where $p(s_1,\vec\kappa_1,s_2,\vec\kappa_2)_{\rm old}$ is the
standard result obtained by means of ordinary quantum optics. 
Taking, as before, $c_n=1/\sqrt{n}$ we find
\be
p(s_1,\vec\kappa_1,s_2,\vec\kappa_2)
&=&\Big(1-\sum_{n=1}^\infty \frac{p_n}{n}\Big) 
|\phi_{s_1,\vec \kappa_1}|^2
|\phi_{s_2,\vec \kappa_2}|^2
p(s_1,\vec\kappa_1,s_2,\vec\kappa_2)_{\rm old}
\ee
Under such assumptions the angular distribution of the two-photon
emission is the same as in the standard theory. The probability of
the 2-photon spontaneous emission is thus a product of four terms. It may
be difficult to distinguish between 
$F(\omega_{\vec\kappa_1})F(\omega_{\vec\kappa_2})
\big(1-\langle\frac{1}{n}\rangle\big)$ and analogous
factors arising in real experiments 
from detector inefficiency. The above result may have therefore
nontrivial implications for the problem of testing quantum mechanics
versus local hidden-variables theories and is very closely related to
the so-called detector inefficiency loophole in Bell's theorem
(cf. \cite{Sven,Gisin}). The reason is that the presence of 
$F(\omega_{\vec\kappa_1})F(\omega_{\vec\kappa_2})
\big(1-\langle\frac{1}{n}\rangle\big)$ will necessarily 
lower two-photon coincidence rates, whereas it is known that 
in order to violate 
the Bell inequality the rates must exceed certain thresholds. 
The problem is worth further studies.

\section{Blackbody radiation}

One of the possible tests of the new formalism is the problem of 
blackbody radiation. Planck's famous formula \cite{Planck}
\be
\varrho(\omega)
=\frac{\hbar}{\pi^2c^3}\frac{\omega^3}{e^{\beta\hbar\omega}-1}
=\frac{\hbar}{\pi^2c^3}\omega^3 \overline{N}_\omega,
\ee
where $\overline{N}_\omega$ is the average number of excitations of
an oscillator in inverse temperature $\beta$,
is one of the first great sucesses of quantum radiation theory and
marks the beginning of quantum mechanics. 
Contemporary measurements of $\varrho(\omega)$ \cite{COBE,COBE99} 
performed by means of COBE (Cosmic Background Explorer) are in a very
good agreement with the Planck law. The data have been carefully
analyzed in the context of nonextensive statistics
\cite{Tsallis1,Tsallis2} in search of possible deviations from
extensivity. The result that comes out systematically is
$|q-1|<10^{-4}$ where $q$ is the Tsallis parameter. The case $q=1$
corresponds to the exact Planck formula. If 
there are any corrections whatever, they must be quite
small. 

The standard derivation of the formula consists basically of two
steps. First, one counts the number of different wave vectors $\vec
k$ such that $c|\vec k|\in [\omega,\omega +\Delta\omega]$. Second,
one associates with each such a vector an oscillator and counts the
average number of its excitations assuming the Boltzmann-Gibbs
probability distribution at temperature $T$ and chemical potential
$\mu=0$. The latter assumption is justified by the fact that the
number of excitations of the electromagnetic field is not conserved in
atom-light interactions. 

In the new model the situation is slightly different since there
exists an additional conserved quantum number: The number of {\it
oscillators\/}. As we have seen in previous calculations the
Hamiltonian is block-diagonal with respect to $\oplus$ but changes
the number of excitations in each $N$-oscillator subspace of the
direct sum. 
The state vectors at the multi-oscillator level are symmetric with
respect to permutations of the oscillators and therefore the
oscillators themselves have to be
regarded as bosons whose number is conserved and their chemical potential is
$\mu\neq 0$. However, their excitations should be regarded as bosons
with vanishing chemical potential. 

The energy eigenvalues 
\be
E_{m,n}=m\hbar\omega\Big(n+\frac{1}{2}\Big)
\ee
corresponding to the oscillator whose frequency
is $\omega$ are parametrized by two natural numbers: $m$ (the number
of oscillators) and $n$ (the number of excitations). Assuming the
standard Boltzmann-Gibbs statistics we obtain the probabilities 
\be
p_{m,n}=Z^{-1}e^{-\beta [m\hbar\omega(n+\frac{1}{2})-m\mu]}
\ee
where 
\be
Z &=& 
\sum_{m=1}^\infty e^{\beta m(\mu+\hbar\omega/2)}
\frac{e^{-\beta m\hbar\omega}}
{1-e^{-\beta m\hbar\omega}}.\label{Z}
\ee
The Lambert series \cite{F} 
\be
\sum_{m=1}^\infty a_m\frac{x^m}{1-x^m}\label{Lambert}
\ee
is convergent for any $x$ if $\sum_{m=1}^\infty a_m$ is convergent. 
Otherwise (\ref{Lambert}) converges for exactly those $x$ for
which the power series $\sum_{m=1}^\infty a_mx^m$ does.
In (\ref{Z}) $a_m=e^{\beta m(\mu+\hbar\omega/2)}$ and 
$\sum_{m=1}^\infty a_m<\infty$ if $\mu+\hbar\omega/2<0$. 
If $\mu+\hbar\omega/2\geq 0$ we still have convergence of (\ref{Z}) 
as long as 
$\sum_{m=1}^\infty
e^{-\beta m[\frac{1}{2}\hbar\omega -\mu]}<\infty$. The upper limit
imposed on $\mu$ by the finiteness of $Z$ is therefore 
$\frac{1}{2}\hbar\omega -\mu>0$. In what follows we assume that 
$\mu$ is 
$\omega$-independent and therefore $\mu\leq 0$.

The appropriate average number of excitations  is
\be
\overline{n}_\omega
&=&
Z^{-1}\sum_{m=1}^\infty\sum_{n=0}^\infty mn 
e^{-\beta [m\hbar\omega(n+\frac{1}{2})-m\mu]}
\ee
and the Planck formula is replaced by
\be
\varrho_{\rm new}(\omega)
&=&
\frac{\hbar}{\pi^2c^3}\omega^3\overline{n}_\omega.
\ee
It is easy to show that $\varrho_{\rm new}(\omega)$ tends to the Planck 
distribution with $\mu\to -\infty$. To see this consider a more general 
series
\be
Z^{-1}\sum_{m=1}^\infty\sum_{n=0}^\infty mn 
q_m e^{-\beta m\hbar\omega(n+\frac{1}{2})}\label{series}
\ee
where $Z$ is the normalization factor and 
$\sum_{m=1}^\infty q_m<\infty$. If $q_1=1$ and $q_m=0$ for $m>1$ 
then (\ref{series}) is just the exact Planckian formula. 
Factoring out $e^{-\beta|\mu|}$ in both the numerator and the denominator 
of $\overline{n}_\omega$ we obtain $q_1=1$ and $q_m=e^{-\beta|\mu|(m-1)}$ 
for $m>1$. For $|\mu|\to \infty$ all $q_m$, for $m>1$, vanish 
and the limiting distribution is Planckian. 

This proves that an experimental agreement with the ordinary Planck's
$\varrho(\omega)$ cannot rule out our modification but can, at most, 
set a lower bound on an admissible value of $|\mu|$.
However, assuming that $\mu$ has 
some finite and fixed value it should be in principle measurable. 
The plots show that the modifications become visible around
$\mu\approx -3k_BT$. Assuming that the chemical potential is
temperature independent, say $\mu=-k_BT_0$, we obtain a kind of
critical temperature $T_{\rm critical}\approx T_0/3$ above which the
ratio $\mu/(k_BT)$ is small enough to make the modifications of the
distribution observable. For $T<T_{\rm critical}$ the distribution
should be given by the Planck law; for $T>T_{\rm critical}$ the
distribution should approach the $\mu=0$ distribution, i.e. this
would be a Planck-type curve but with the maximum lowered and shifted
towards higher energies.

Fig.~1 shows the plots of $\varrho_{\rm new}(\omega)$ for $\mu=0$
(lower dotted),
$\mu= -0.8 k_B T$ (upper dotted), and $\mu= -10 k_B T$ (solid). 
The thick dashed curve is the Planck distribution. The curve obtained
for $\mu=-10 k_B T$ is indistinguishable from the Planck
distribution. 
The plot does not change if one takes $\mu< -10
k_B T$ and differences are not visible even if one plots the distributions
in logarithmic scales (not shown here).
This is a numerical proof that the distribution we have
obtained on the basis of the modified quantization tends very quickly
to the Planck one as $\mu\to -\infty$. 
It is instructive to compare the modification we have predicted with
those arising from nonextensive statistics. 
The two thin dashed lines represent Tsallis distributions resulting
from nonextensive formalism for $q=0.95$ (lower) and $q=1.05$ (upper). 
The modifications we have derived are therefore qualitatively different from
those resulting from Tsallis statistics. 
\begin{figure}
\epsfxsize=4in \epsfbox{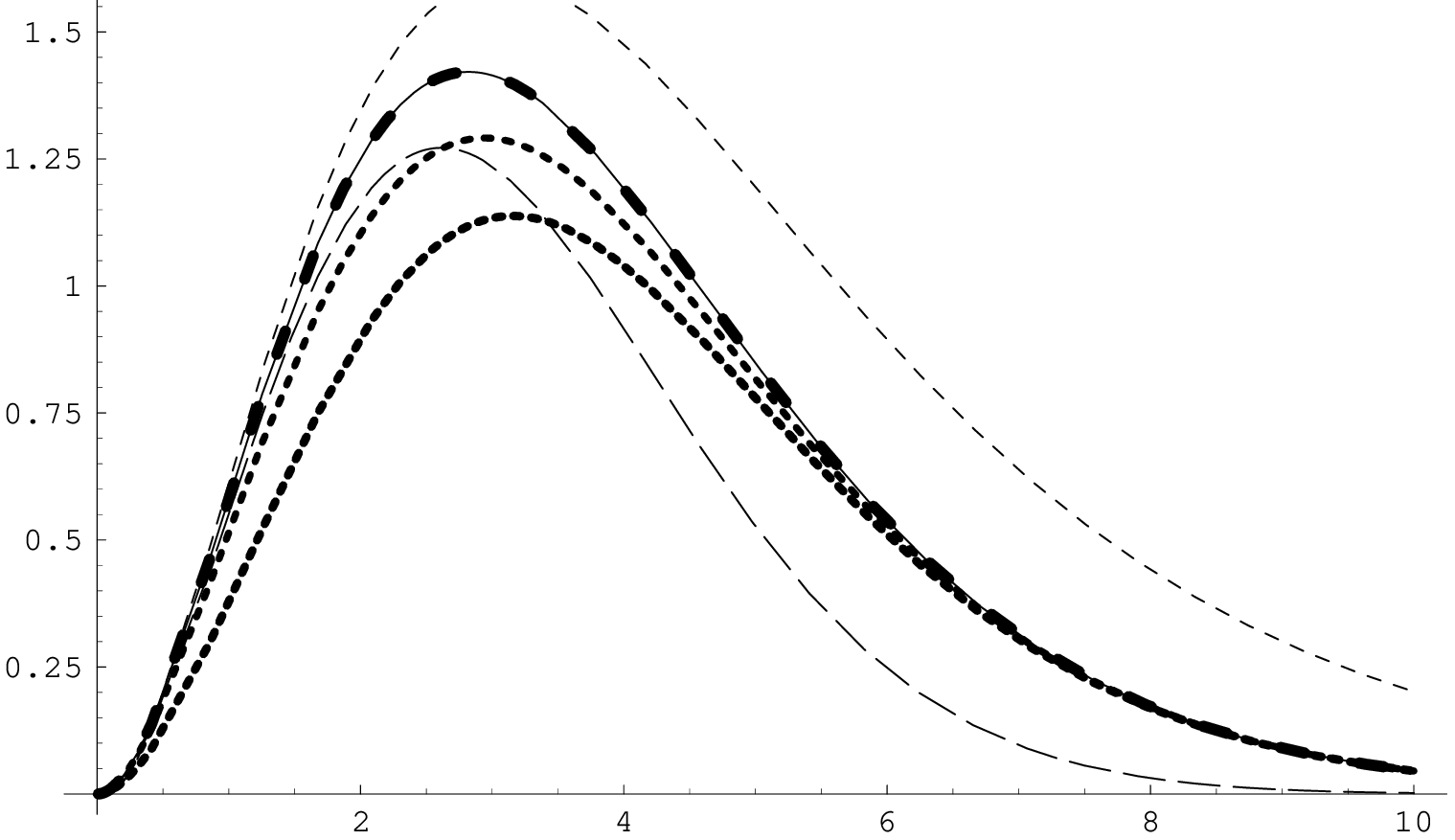}
\caption{
$\varrho_{\rm new}(\omega)$ for $\mu=0$
(lower dotted),
$\mu= -0.8 k_B T$ (upper dotted), and $\mu= -10 k_B T$ (solid). The 
energy range is $0.01 k_B T<\hbar\omega <10 k_B T$.
The thick dashed curve is the Planck distribution. The curve obtained
for $\mu=-10 k_B T$ is indistinguishable from the Planck
distribution.
The two thin dashed lines represent Tsallis distributions resulting
from the Tsallis formalism for $q=0.95$ (lower) and $q=1.05$ (upper).
Since $\varrho_{\rm new}<\varrho$ at least in the neighborhood of the
maximum, the new distribution has to be compared with $q<1$
statistics. The curves are qualitatively different. In particular,
all $q<1$ distributions require an energy cut-off which does not
occur for $\varrho_{\rm new}$.}
\end{figure}
\begin{figure}
\epsfxsize=4in \epsfbox{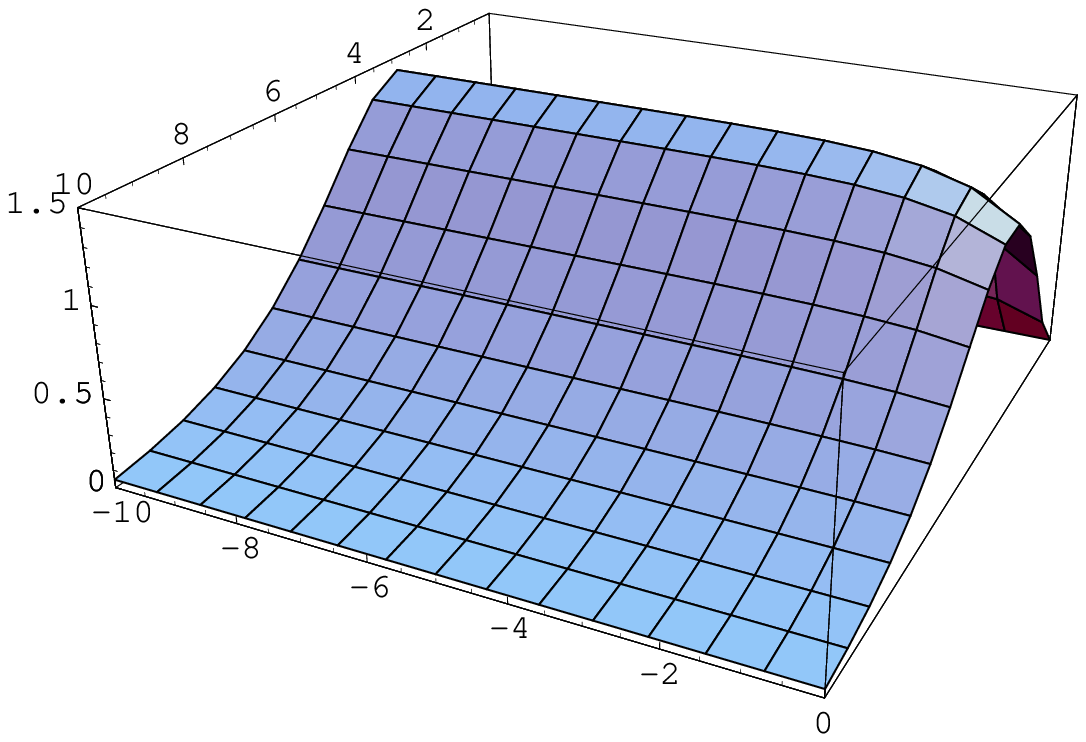}
\caption{
$\varrho_{\rm new}(\omega)$ for $-10k_BT\leq \mu\leq 0$. The cut
through $\mu=-10 k_BT$ is practically indistinguishable from the Planck
distribution.}
\end{figure}
\begin{figure}
\epsfxsize=4in \epsfbox{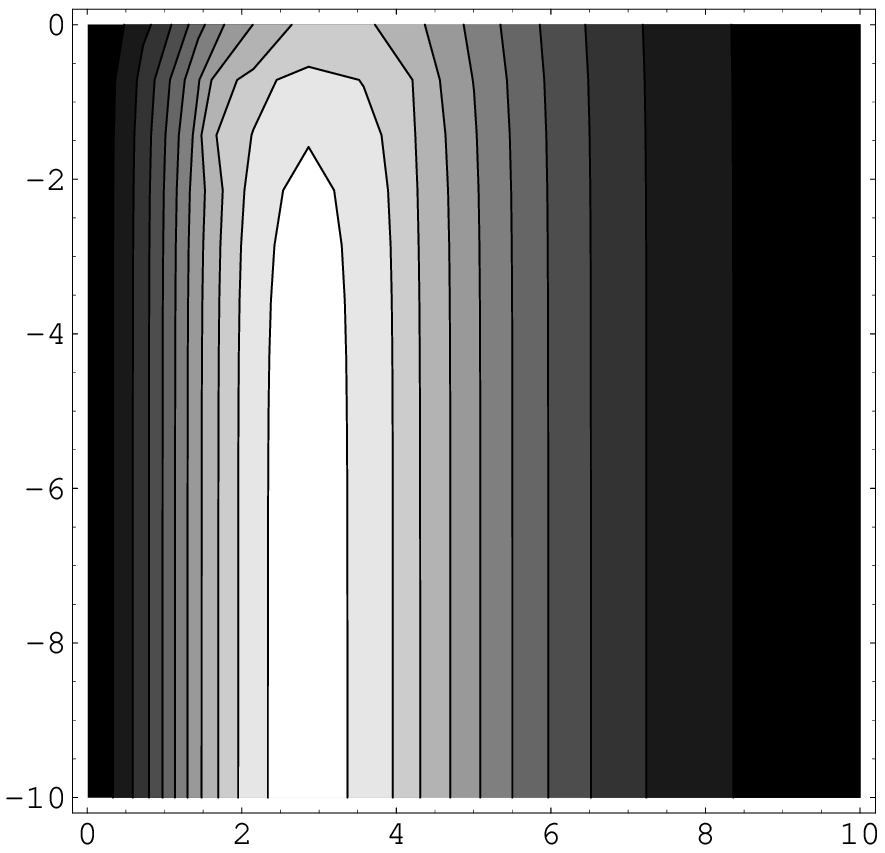}
\caption{
Contour plot of $\varrho_{\rm new}(\omega)$ for $-10k_BT\leq \mu\leq
0$. The fast convergence to Planckian $\varrho(\omega)$ (as $\mu\to
-\infty$) is clearly seen. 
}
\end{figure}

\section{Conclusions}

``A theory that is as spectacularly successful as quantum
electrodynamics has to be more or less correct, although we may not
be formulating it in just the right way" \cite{Dreams}. The above
quotation from Weinberg could serve as a motto opening our paper. The
main idea we have tried to advocate was that the standard canonical
quantization procedure is, in certain sense, {\it too classical\/} to
be good. 

The reasons for such a choice of quantization  
could be both historical and sociological and may be
rooted in the fact that the idea of quantizing the field was
formulated before the real development of modern quantum mechanics. 
In oscillations of a simple pendulum it may be justified to treat
$\omega$ as an external parameter defining the system (via, say, the
length of the pendulum). But oscillations of the electromagnetic
field do not seem to have such a ``mechanical" origin 
and it is more natural to think of the
spectrum of frequencies as eigenvalues of some Hamiltonian. That is
exactly what happens with other quantum wave equations. 

We have defined the quantum electromagnetic field as an oscillator
that can exist in a superposition of different frequencies (or,
rather, wave vectors). This should not be confused with the classical
superpositions of frequencies created by, say, a guitar string. The
superpositions we have in mind dissappear at the classical level. 

Once one accepts this viewpoint it becomes clear how to quantize the
field at the level of a single oscillator. We do not need many
oscillators to perform the field quantization. But 
there is no reason to believe that all the possible fields can be
described by the same single oscillator. And even more: We know that
the structure of the one-oscillator Hilbert space is not rich enough to
describe multi-particle entangled states and there is no doubt that
such states are physical. The next step, performed already {\it after\/}
the quantization, is to consider fields consisting of 1, 2, 3 and more
oscillators, and even existing in superpositions of different numbers
of them. The resulting structure is analogous to the Fock
space but so the procedure can be 
(although somewhat misleadingly) referred to as ``second quantization''.  
What is essential we do not need the vacuum state understood as
the cyclic vector of the GNS construction. 
In the new framework such an object seems 
rather artificial. 

On the other hand, there exist vacuum {\it states\/}. These are all
the states describing ground states of the oscillators. 
They correspond to concrete {\it finite\/} average values of energy. A
general vacuum state is therefore a superposition of different
eigenstates of a free Hamiltonian and is not, in itself, an
eigenstate of the Hamiltonian. 

No further assumptions are made. The system is described by laws of
ordinary quantum mechanics so that to compute concrete problems we
can use standard methods. Perturbation theory
leads to structures we know from the standard Feynmann diagrams.
The blackbody radiation is calculated by means of the standard
Boltzmann-Gibbs statistics. 

Let us close these remarks with another quotation: ``Present quantum 
electrodynamics contains many very important `elements of truth', but also 
some clear `elements of nonsense'. Because of the divergences and ambiguities, 
there is general agreement that a rather deep modification of the theory 
is needed, but in some forty years of theoretical work, nobody has seen how to 
disentengle the truth from the nonsense. In such a situation, one needs 
more experimental evidence, but during that same forty years we have found 
no clues from the laboratory as to what specific features of QED might be
modified. Even worse, in the absence of any alternative theory whose 
predictions differ from those of QED in known ways, we have no criterion 
telling us {\it which\/} experiments would be relevant ones to try. 

It seems useful, then, to examine the various disturbing features of QED, 
which give rise to mathematical or conceptual difficulties, to ask
whether present empirical evidence demands their presence, and to explore 
the consequences of the modified (although perhaps rather crude and 
incomplete) theories in which these features are removed. Any difference 
between the predictions of QED and some alternative theory, corresponds 
to an experiment which might distinguish between them; if it appears untried
but feasible, then we have the opportunity to subject QED to a new test 
in which we know just what to look for, and which we would be very unlikely 
to think of without the alternative theory. For this purpose, the alternative
theory need not be worked out as completely as QED; it is sufficient 
if we know in what way their predictions will differ in the area of 
interest. Nor does the alternative theory need to be free of defects in 
all other respects; for if experiment should show that it contains just a 
single `element of truth' that is {\it not\/} in QED, then the alternative 
theory will have served its purpose; we would have the long-missing clue
showing in what way QED must be modified, and electrodynamics (and, I suspect,
much more of theoretical physics along with it) could get moving again
`` \cite{Janes}. 
\acknowledgements

This work was done partly during my stays in Arnold Sommerfeld
Institute in Clausthal. I gratefully acknowledge a support from the
Alexander von Humboldt Foundation. I'm indebted to Prof. Iwo
Bia{\l}ynicki-Birula, Robert~Alicki, Jan Naudts, and Wolfgang Luecke 
for critical comments, and Pawe{\l} Syty for a stimulating discussion 
on small $\omega$'s. 

\section{Appendix: Technical differences and similarities with
respect to the standard formalism}

\subsection{Proof of Eq.~(11)}
\label{A-imp}

This calculation is elementary but very important, so we give it explicitly:
\be
a_{\omega_k}(t)
&=&
e^{iHt/\hbar}a_{\omega_k}e^{-iHt/\hbar}\\
&=&
e^{i\Omega\otimes \big(a^{\dag}a+\frac{1}{2}\bbox 1\big)t}
|\omega_k\rangle\langle \omega_k| 
\otimes a
e^{-i\Omega\otimes \big(a^{\dag}a+\frac{1}{2}\bbox 1\big) t}\\
&=&
|\omega_k\rangle\langle \omega_k|
\otimes
e^{i\omega_k  \big(a^{\dag}a+\frac{1}{2}\bbox 1\big) t} a
e^{-i\omega_k \big(a^{\dag}a+\frac{1}{2}\bbox 1\big) t}\\
&=&
|\omega_k\rangle\langle \omega_k|
\otimes
e^{-i\omega_k t} a
=
e^{-i\omega_k t} a_{\omega_k}(0)
\ee

\subsection{Energy-momentum operators for free fields: 1-oscillator
formulas} 
\label{A-EP}
To see how (\ref{P1}) and (\ref{P2}) arise 
let us first note that  
\be
a_{s_1,\vec \kappa_1}a_{s_2,\vec \kappa_2}&=&0\\
a_{s_1,\vec \kappa_1}a^{\dag}_{s_2,\vec \kappa_2}&=&0\\
a^{\dag}_{s_1,\vec \kappa_1}a^{\dag}_{s_2,\vec \kappa_2}&=&0
\ee
unless $s_1=s_2$ and $\vec \kappa_1=\vec \kappa_2$ [cf. Eqs.
(\ref{a}) and (\ref{a^dag})]. The terms involving
$(a_{s,\vec \kappa})^2$ and $(a^{\dag}_{s,\vec \kappa})^2$ disappear
due to $\vec e_{s,\vec \kappa}\cdot \vec e_{s,\vec \kappa}
=0$ and its complex conjugate. As a result
\be
\frac{1}{2}
\Big(
\hat{\vec E}(t,\vec x)\cdot \hat{\vec E}(t,\vec x)
+
\hat{\vec B}(t,\vec x)\cdot \hat{\vec B}(t,\vec x)
\Big)
&=&
e^{i{P}\cdot x/\hbar}
\sum_{s,\kappa_\lambda }
\frac{\hbar\omega_\lambda}{2V}
|s,\kappa_\lambda\rangle\langle s,\kappa_\lambda|\otimes 
\big(
a a^{\dag}+ a^{\dag} a
\big)
e^{-i{P}\cdot x/\hbar}\nonumber\\
&=&
\sum_{s,\kappa_\lambda }
\frac{\hbar\omega_\lambda}{2V}
|s,\kappa_\lambda\rangle\langle s,\kappa_\lambda|\otimes 
\big(
a a^{\dag}+ a^{\dag} a
\big)=H/V.
\nonumber
\ee
To find the relation between the Pointing vector and linear momentum
we first have to show that 
\be
\hat{\vec E}\times \hat{\vec B}
=- \hat{\vec B}\times \hat{\vec E}.
\ee
The relevant formula is 
\be
{[\hat E_\alpha,\hat B_\beta]}
&=&
\sum_{s,\kappa_\lambda }
i\hbar\omega_\lambda
\frac{s}{2}
\Big(
\delta_{\alpha\beta}
-
(n_{\vec \kappa_{\lambda}})_\alpha
(n_{\vec \kappa_{\lambda}})_\beta
\Big)
|s,\kappa_\lambda\rangle\langle s,\kappa_\lambda|\otimes
\bbox 1.
\ee
The remaining calculations are similar to those for $H$.

\subsection{Derivation of the ``2-photon" amplitude: 2 oscillators}
\label{A-2in2}
We employ the standard second-order time dependent perturbation theory
and notation from \cite{C-T}. 
\be
{}&{}&
c_2^{-2}
\langle b|\langle s_1,\vec \kappa_1,1|
\langle s_2,\vec \kappa_2,1|U^{(2)}(t_f,t_i)|a\rangle
|\phi\rangle|\phi\rangle
\nonumber\\
&{}&=
-\frac{2\pi ie^2}{m^2}
\sum_{c,S_1,S_2,\vec K_1,\vec K_2,n_1,n_2}
\sum_{r,\vec k,r',\vec k'}\phi_{r,\vec k}\phi_{r',\vec k'}
\frac{\delta^{(T)}(E_{a,\vec \kappa_1,0,\vec \kappa_2,0}-
E_{b,\vec \kappa_1,1,\vec \kappa_2,1})}
{E_{a,\vec k,0,\vec k,0'}-E_{c,\vec K_1,n_1,\vec K_2,n_2}+i0_+}
\nonumber\\
&{}&\pp =\times
\langle b|\langle s_1,\vec \kappa_1,1|
\langle s_2,\vec \kappa_2,1|
\sum_{s,\vec\kappa}
\sqrt{\frac{\hbar}{2\omega_{\vec\kappa} V}}
\Big(\bbox a_{s,\vec\kappa}
\vec e_{s,\vec\kappa}\cdot\vec p
+
\bbox a^{\dag}_{s,\vec\kappa}
\vec e^{\,*}_{s,\vec\kappa}\cdot\vec p
\Big)
|c\rangle
|S_1,\vec K_1,n_1\rangle|S_2,\vec K_2,n_2\rangle\nonumber\\
&{}&\pp =\times
\langle c|\langle S_1,\vec K_1,n_1|\langle S_2,\vec K_2,n_2|
\sum_{s',\vec\kappa'}
\sqrt{\frac{\hbar}{2\omega_{\vec\kappa'} V}}
\Big(\bbox a_{s',\vec \kappa'}
\vec e_{s',\vec\kappa'}\cdot\vec p
+
\bbox a^{\dag}_{s',\vec\kappa'}
\vec e^{\,*}_{s',\vec\kappa'}\cdot\vec p
\Big)
|a\rangle
|r,\vec k,0\rangle|r',\vec k',0\rangle
\nonumber\\
&{}&=
-\frac{2\pi ie^2}{m^2}
\sum_{c,S_1,S_2,\vec K_1,\vec K_2,n_1,n_2}
\sum_{r,\vec k,r',\vec k'}\phi_{r,\vec k}\phi_{r',\vec k'}
\frac{\delta^{(T)}(E_{a,\vec \kappa_1,0,\vec \kappa_2,0}-
E_{b,\vec \kappa_1,1,\vec \kappa_2,1})}
{E_{a,\vec k,0,\vec k',0}-E_{c,\vec K_1,n_1,\vec K_2,n_2}+i0_+}
\nonumber\\
&{}&\pp =\times
\langle s_1,\vec \kappa_1,1|
\langle s_2,\vec \kappa_2,1|
\sum_{s,\vec\kappa}
\sqrt{\frac{\hbar}{2\omega_{\vec\kappa} V}}
\Big(\big(a_{s,\vec\kappa}\otimes \bbox 1
+
\bbox 1\otimes a_{s,\vec\kappa}\big)
\vec e_{s,\vec\kappa}\cdot\vec p_{bc}
+
\big(a^{\dag}_{s,\vec\kappa}\otimes \bbox 1
+
\bbox 1\otimes a^{\dag}_{s,\vec\kappa}\big)
\vec e^{\,*}_{s,\vec\kappa}\cdot\vec p_{bc}
\Big)\nonumber\\
&{}&\pp =\times
|S_1,\vec K_1,n_1\rangle|S_2,\vec K_2,n_2\rangle
\langle S_1,\vec K_1,n_1|\langle S_2,\vec K_2,n_2|\nonumber\\
&{}&\pp =\times
\sum_{s',\vec\kappa'}
\sqrt{\frac{\hbar}{2\omega_{\vec\kappa'} V}}
\Big(
\big(a_{s',\vec\kappa'}\otimes \bbox 1
+
\bbox 1\otimes a_{s',\vec\kappa'}\big)
\vec e_{s',\vec\kappa'}\cdot\vec p_{ca}
+
\big(a^{\dag}_{s',\vec\kappa'}\otimes \bbox 1
+
\bbox 1\otimes a^{\dag}_{s',\vec\kappa'}\big)
\vec e^{\,*}_{s',\vec\kappa'}\cdot\vec p_{ca}
\Big)
|r,\vec k,0\rangle|r',\vec k',0\rangle\nonumber.
\ee
The block-diagonal property of the interaction
Hamiltonian has been used twice. The remaining calculations are standard. 
It is remarkable that although the result we obtain is essentially
the same as in the standard formalism, the technical reasons for this
are completely different. 

\subsection{Derivation of the ``2-photon" amplitude: 3 oscillators}
\label{A-2in3}

Here we sketch the proof of the 3-oscillator amplitude. 
In second-order perturbation theory 
\be
{}&{}&
c_3^{-2}
\langle b|\langle s_1,\vec \kappa_1,1|
\langle s_2,\vec \kappa_2,1|\langle\phi|U^{(2)}(t_f,t_i)|a\rangle|\phi\rangle
|\phi\rangle|\phi\rangle\nonumber\\
&{}&
=
-\frac{2\pi ie^2}{m^2}
\sum_{c,S_1,S_2,\vec K_1,\vec K_2,\vec K_3,n_1,n_2,n_3}
\sum_{r_0,\vec k_0,r,\vec k,r',\vec k',r'',\vec k''}
\phi^*_{r_0,\vec k_0}\phi_{r,\vec k}\phi_{r',\vec k'}\phi_{r'',\vec k''}
\nonumber\\
&{}&\pp =\times
\langle b|\langle s_1,\vec \kappa_1,1|
\langle s_2,\vec \kappa_2,1|\langle r_0,\vec k_0,0|
\sum_{s,\vec\kappa}
\sqrt{\frac{\hbar}{2\omega_{\vec\kappa} V}}
\Big(\bbox a_{s,\vec\kappa}
\vec e_{s,\vec\kappa}\cdot\vec p
+
\bbox a^{\dag}_{s,\vec\kappa}
\vec e^{\,*}_{s,\vec\kappa}\cdot\vec p
\Big)
|c\rangle
|S_1,\vec K_1,n_1\rangle|S_2,\vec K_2,n_2\rangle
|S_3,\vec K_3,n_3\rangle\nonumber\\
&{}&\pp =\times
\langle c|\langle S_1,\vec K_1,n_1|\langle S_2,\vec K_2,n_2|
\langle S_3,\vec K_3,n_3|
\sum_{s',\vec\kappa'}
\sqrt{\frac{\hbar}{2\omega_{\vec\kappa'} V}}
\Big(\bbox a_{s',\vec \kappa'}
\vec e_{s',\vec\kappa'}\cdot\vec p
+
\bbox a^{\dag}_{s',\vec\kappa'}
\vec e^{\,*}_{s',\vec\kappa'}\cdot\vec p
\Big)
|a\rangle
|r,\vec k,0\rangle|r',\vec k',0\rangle|r'',\vec k'',0\rangle
\nonumber\\
&{}&\pp =\times
\frac{\delta^{(T)}(E_{a,\vec k,0,\vec k',0,\vec k'',0}
-E_{b,\vec \kappa_1,1,\vec \kappa_2,1,\vec k_0,0})}
{E_{a,\vec k,0,\vec k',0,\vec k'',0}
-E_{c,\vec K_1,n_1,\vec K_2,n_2,\vec K_3,n_3}+i0_+}
\nonumber\\
&{}&
=
-\frac{2\pi ie^2}{m^2}
\sum_{c,S_1,S_2,\vec K_1,\vec K_2,\vec K_3,n_1,n_2,n_3}
\sum_{r_0,\vec k_0,r,\vec k,r',\vec k',r'',\vec k''}
\phi^*_{r_0,\vec k_0}\phi_{r,\vec k}\phi_{r',\vec k'}\phi_{r'',\vec k''}
\nonumber\\
&{}&\pp =\times
\langle b|\langle s_1,\vec \kappa_1,1|
\langle s_2,\vec \kappa_2,1|\langle r_0,\vec k_0,0|
\sum_{s,\vec\kappa}
\sqrt{\frac{\hbar}{2\omega_{\vec\kappa} V}}
\Big(\bbox a_{s,\vec\kappa}
\vec e_{s,\vec\kappa}\cdot\vec p
+
\bbox a^{\dag}_{s,\vec\kappa}
\vec e^{\,*}_{s,\vec\kappa}\cdot\vec p
\Big)\nonumber
\ee
\be
{}&{}&\pp =\times
|c\rangle
|S_1,\vec K_1,n_1\rangle|S_2,\vec K_2,n_2\rangle
|S_3,\vec K_3,n_3\rangle
\langle c|\langle S_1,\vec K_1,n_1|\langle S_2,\vec K_2,n_2|
\langle S_3,\vec K_3,n_3|\nonumber\\
{}&{}&\pp =\times
\sum_{s',\vec\kappa'}
\sqrt{\frac{\hbar}{2\omega_{\vec\kappa'} V}}
\Big(
a^{\dag}_{s',\vec\kappa'}\otimes \bbox 1 \otimes \bbox 1
+
\bbox 1\otimes a^{\dag}_{s',\vec\kappa'}\otimes \bbox 1
+
\bbox 1 \otimes \bbox 1\otimes a^{\dag}_{s',\vec\kappa'}
\Big)\vec e^{\,*}_{s',\vec\kappa'}\cdot\vec p
|a\rangle
|r,\vec k,0\rangle|r',\vec k',0\rangle|r'',\vec k'',0\rangle
\nonumber\\
&{}&\pp =\times
\frac{\delta^{(T)}(E_{a,\vec k,0,\vec k',0,\vec k'',0}
-E_{b,\vec \kappa_1,1,\vec \kappa_2,1,\vec k_0,0})}
{E_{a,\vec k,0,\vec k',0,\vec k'',0}
-E_{c,\vec K_1,n_1,\vec K_2,n_2,\vec K_3,n_3}+i0_+}
\nonumber
\ee
The remaining part of the proof is standard. In the course of the
computation one recognizes the elements known from standard Feynman
diagrams, in particular the self-energy corrections due to emission
and reabsorbtion of virtual photons. A general property of the
perturbation series we find is its better convergence due to the
presence of the vacuum amplitudes $\phi_{s,\vec k}$.

\end{document}